\documentclass[a4paper,11pt]{article}    
\pdfoutput=1 
\usepackage{jheppub}
\usepackage{epsfig}
\usepackage[T1]{fontenc} 
\def\arXiv#1{\href{http://arxiv.org/abs/#1}{arXiv:#1}}
\def\arXiv#1#2{\href{http://arxiv.org/abs/#1}{arXiv:#1}}

\def\doi#1#2{\href{http://doi.org/#1}{#2}}

\def\be{\begin{eqnarray}}
\def\ee{\end{eqnarray}}
\def\bea{\begin{eqnarray}}
\def\eea{\end{eqnarray}}
\newcommand{\nn}{\nonumber}
\newcommand\para{\paragraph{}}

\def\Dslash{\,\,{\raise.15ex\hbox{/}\mkern-12mu D}}
\def\Dbarslash{\,\,{\raise.15ex\hbox{/}\mkern-12mu {\bar D}}}
\def\delslash{\,\,{\raise.15ex\hbox{/}\mkern-9mu \partial}}
\def\delbarslash{\,\,{\raise.15ex\hbox{/}\mkern-9mu {\bar\partial}}}
\def\pslash{\,\,{\raise.15ex\hbox{/}\mkern-9mu p}}
\def\calDslash{\,\,{\raise.15ex\hbox{/}\mkern-12mu {\cal D}}}

\def\lae{\mathrel{\mathop{\smash{\lower .5 ex \hbox{$\stackrel<\sim$}}}}}
\def\lae{\mathrel{\mathop{\smash{\lower .5 ex \hbox{$\stackrel>\sim$}}}}}

\title{\boldmath  Weyl semimetal/insulator transition from holography}

\author{Yan Liu and Junkun Zhao}
\affiliation{Department of Space Science, and International Research Institute
of Multidisciplinary Science,
\\ Beihang University,  Beijing 100191, China}
\emailAdd{yanliu@buaa.edu.cn,  junkunzhao@buaa.edu.cn}

\abstract{
We study a holographic model which exhibits a quantum phase transition from the strongly interacting Weyl semimetal phase to an insulating phase. In the holographic insulating phase there is a hard gap in the real part of frequency dependent diagonal conductivities. However, the anomalous Hall conductivity is nonzero at zero frequency, indicting that it is a Chern insulator. 
This holographic quantum phase transition is always of first order, signified by a discontinuous anomalous Hall conductivity at the phase transition, in contrast to the very continuous holographic Weyl semimetal/trivial semimetal phase transition. 
Our work reveals the novel phase structure of strongly interacting Weyl semimetal.}

\begin{document}
\maketitle
\flushbottom
\pagestyle{plain} \setcounter{page}{1}
\newcounter{bean}
\baselineskip16pt



\section{Introduction}

Weyl semimetal is a nontrivial topological gapless state and exhibits lots of exotic novel and robust properties, including chiral anomaly and Fermi arc etc. It has been a research focus recently as they are both experimentally important and theoretically interesting \cite{Witten:2015aoa, vishwanath,burkov0}. On the one hand, they are ideal systems to test the 
macroscopic effects due to quantum anomaly \cite{Landsteiner:2016led}, including chiral magnetic effect and transport effects induced by the mixed axial-gravitational anomaly. 
On the other hand, Weyl semimetal is a novel kind of topological quantum matter which goes beyond the Landau-Ginzburg's paradigm for classification of states of matter. 
Similar to graphene systems \cite{jan}, Weyl semimetal systems could be strongly coupled and do not possess well-defined 
quasiparticles. It is theoretically challenging and important to study strongly interacting Weyl semimetals, to go beyond the conventional approach on topological states of matter based on topological band theory or weakly coupled theory.  

\para
The holographic correspondence maps the difficult question of strongly interacting field theory to a tractable weakly coupled gravitational problem. There have been lots of remarkable applications of holography to tackle the strongly interacting condensed matter questions \cite{Zaanen:2015oix,book0,review}. In particular, 
holographic models for strongly interacting Weyl semimetals have been constructed in \cite{Landsteiner:2015pdh, Landsteiner:2015lsa} in which the anomalous Hall conductivity is an order parameter to characterize the quantum topological phase transition. The effects of the surface state \cite{Ammon:2016mwa} and topological invariants \cite{Liu:2018djq} in this holographic model exhibit key features of topological Weyl semimetals. Therefore with strong interaction topological Weyl semimetal still exits and holography is a practical tool to explore its property. Moreover, the nontrivial topological structure in the strongly interacting system can be revealed from the gravitational bulk physics \cite{Liu:2018bye, Liu:2018djq}.  There exist two bulk matter fields in which one field is to generate a gap in the dual theory while the other matter field is to deform the Fermi points to a topologically nontrivial  configuration  (Weyl points or nodal lines). The different topological phases arise due to the different IR solutions in the bulk  which are adiabatically disconnected and only one of the matters fields dominates in each solution. From the holographic model a nontrivial prediction is that the presence of odd viscosity is due to mixed axial-gravitational anomaly  \cite{Landsteiner:2016stv}. Other various interesting aspects of holographic Weyl semimetals have been explored, including optical conductivity  \cite{Grignani:2016wyz}, axial Hall conductivity \cite{Copetti:2016ewq}, disorder effect on topological phase transition  \cite{Ammon:2018wzb} and  the butterfly velocity \cite{Baggioli:2018afg}.

\para
In condensed matter systems, from weakly coupled theory Weyl semimetal can go through a quantum phase transition to a normal band insulator \cite{burkov1, Roy:2016rqw} or to Chern insulator \cite{burkov1,cm-1,Roy:2016amv} etc.\footnote{See appendix \ref{app:a} for examples from field theoretical approach.} It would be extremely interesting to explore the phase diagram of strongly interacting topological Weyl semimetal from holography. In the previous holographic models \cite{Landsteiner:2015pdh, Landsteiner:2015lsa} only a portion of degrees of freedom are gapped in the trivial phase and Weyl semimetal phase goes to a trivial semimetal phase after the phase transition. This paper aims to provide a holographic model to describe a quantum phase transition from Weyl semimetal to a phase in which all the degrees of freedom are gapped, namely, the trivial phase is instead an insulating phase. In doing so we start from the most generic holographic Weyl semimetal model by using the Stueckelberg trick to replace the complex scalar field  in \cite{Landsteiner:2015pdh} by two real scalar fields and introduce the most general dilatonic coupling. Writing the equations for fluctuations of gauge fields into a Schrodinger equation, we can get the condition for the dilatonic couplings to produce the insulating phase. 
With a proper choice of dilatonic coupling and potential terms, we could realise a holographic topological quantum phase transition between strongly interacting Weyl semimetal phase and Chern insulator (3+1D anomalous Hall state) phase.  Then we show the evidences of the phase transition from the perspectives of free energy and conductivities. 

\para
Our paper is organized as follows. In section \ref{sec2}, we introduce a generalized holographic model with dilatonic coupling to realise the quantum phase transition from Weyl semimetal phase to insulator phase and show that it is a first order phase transition. In section \ref{sec3}, the conductivities of the dual theory are explored by studying the vector gauge field fluctuations above the background geometry. Evidence for the insulating phase being a Chern insulator is discussed. In section \ref{sec4}, we conclude and discuss the open problems. Appendices \ref{app:a}, \ref{app:b}, \ref{app:c} are devoted to the details of the field theory model, the finite temperature equations of motion for holographic model and the Schrodinger potential approach for conductivities. 

\section{Holographic setup}
\label{sec2}
\para
We shall start from the most general holographic system which duals to an anomalous system with $U(1)_V\times U(1)_A$. This $U(1)_A$ will be explicitly broken by turning on a source term which plays the similar role of mass effect in the dual field theory.  
With the dilatonic coupling,
the generic holographic model is
\bea
\mathcal{S}&=&\int d^5x\sqrt{-g}\bigg[\frac{1}{2\kappa^2}\big(R+12\big)-\frac{Y(\phi)}{4}\mathcal{F}^2-\frac{Z(\phi)}{4}F^2+
\frac{\alpha}{3}\epsilon^{abcde}A_a\Big(F_{bc}F_{de}+3\mathcal{F}_{bc}\mathcal{F}_{de}\Big)\nn\\
&&~~~~~~~~~~~~\label{eq:action}
-\frac{1}{2}(\partial\phi)^2-\frac{W(\phi)}{2}(A_a-\partial_a\theta)^2-V(\phi)
\bigg]
\eea
with the vector gauge field strenght
$\mathcal{F}_{ab}=\partial_a V_b-\partial_b V_a$ and axial gauge field strength $F_{ab}=\partial_a A_b-\partial_b A_a.
$  Here $V_a$ and $A_a$ correspond to vector and axial current respectively. Note that the scalar fields $\phi$ and $\theta$ are real which are dual to operators $\bar{\psi}\psi$ and $\bar{\psi}\gamma^5\psi$.  The action is invariant under the gauge transformation $\theta\to \theta+\chi, ~A_a\to A_a+\partial_a\chi$. One can recover the holographic Weyl semimetal model in \cite{Landsteiner:2015pdh} via $\Phi=\frac{1}{\sqrt{2}}\phi e^{i\theta}$, and choose $Y(\phi)=Z(\phi)=1, ~W(\phi)=q^2\phi^2, ~ V(\phi)=\frac{m^2}{2}\phi^2$. The model (\ref{eq:action}) is a generic holographic model for Weyl semimetal.\footnote{Note that
$\epsilon_{abcde}=\sqrt{-g}\varepsilon_{abcde}$ with
$\varepsilon_{0123r}=1$. Similar generalisation has been made in \cite{Grignani:2016wyz} to study the optical conductivity in the quantum critical regime.
}
\para
From now on we set $2\kappa^2=1.$
The equations of motion of the system are
\bea
R_{ab}-\frac{1}{2}g_{ab}\big(R+12\big)-\frac{1}{2}T_{ab}&=&0 \,,\nn\\
\nabla_b\big(Y(\phi) \mathcal{F}^{ba}\big)+2\alpha\epsilon^{abcde}F_{bc}\mathcal{F}_{de}&=&0 \,,\nn\\
\nabla_b\big(Z(\phi) F^{ba}\big)+\alpha\epsilon^{abcde}\big(F_{bc}F_{de}
  +\mathcal{F}_{bc}\mathcal{F}_{de}\big)-W(\phi) (A^a-\nabla^a \theta)&=&0\,,\nn\\
\nabla_{a}\nabla^{a}\phi-\frac{\partial_\phi Y(\phi)}{4}\mathcal{F}^2-\frac{\partial_\phi Z(\phi)}{4}F^2
  -\frac{\partial_\phi W(\phi)}{2}(A_a-\partial_a\theta)^2-\partial_\phi V(\phi)
&=&0 \,,\nn\\
\nabla_a\left[W(\phi)(A^a-\nabla^a\theta)\right]&=&0 \,,\nn
\eea
where
\bea
T_{ab}&=& Y(\phi)\bigg[\mathcal{F}_{ac}\mathcal{F}_{b}^{~c}-\frac{1}{4}g_{ab}\mathcal{F}^2\bigg]+Z(\phi) \bigg[F_{ac}{F}_{b}^{~c}-\frac{1}{4}g_{ab} F^2\bigg]+W(\phi)\bigg[ (A_a-\partial_a\theta) (A_b-\partial_b\theta)
\nn\\
&&~~~~~~
-\frac{1}{2}g_{ab}  (A_c-\partial_c\theta)^2\bigg]+
\nabla_{a}\phi\nabla_{b}\phi-\frac{1}{2}g_{ab}(\partial\phi)^2-g_{ab}V(\phi)
\,.\nn
\eea

\para
The dual consistent currents can be obtained through the variation of the on-shell action with respect to the gauge fields,
\bea
J^\mu&=&\lim_{r_c\to\infty}\Big[\sqrt{-g}Y \mathcal{F}^{\mu r}+4\alpha\sqrt{-g} \varepsilon^{r\mu\nu\rho\lambda}A_\nu \mathcal{F}_{\rho\lambda}+\frac{\delta S_\text{c.t.}}{\delta v_\mu}\Big]\,,\\
J_5^\mu&=&\lim_{r_c\to\infty}\Big[\sqrt{-g}Z F^{\mu r}+\frac{4}{3}\alpha\sqrt{-g} \varepsilon^{r\mu\nu\rho\lambda}A_\nu F_{\rho\lambda}+\frac{\delta S_\text{c.t.}}{\delta a_\mu}\Big]\,,
\eea
with
\bea
\nabla_\mu J^\mu&=&0\,,\\
\nabla_\mu J_5^\mu&=&\lim_{r_c\to\infty}\Big[\sqrt{-g}W\left(A^r-\nabla^r\theta\right)-
  \frac{\alpha}{3}\sqrt{-g}\varepsilon^{r\mu\nu\rho\lambda}
  (F_{\mu\nu}F_{\rho\lambda}+3\mathcal{F}_{\mu\nu}\mathcal{F}_{\rho\lambda})\Big]+\text{c.t.}\,.
\eea
For simplicity the countertem part is not shown here. Note that the above equations are the dual Ward identities at the operator level. One can always choose the radial gauge $A_r=0$. For a particular state of the dual field theory, i.e. the fluctuation state around the background in the bulk, the term $-\sqrt{-g}Wg^{rr}\partial_r\theta$ plays the role of the explicit breaking term as in the weakly coupled theory which can be found in appendix \ref{app:a}. Since the Ward identity of conserved currents should not depend on the coupling constant of the system, it is expected that this  holographic model describes a strongly interacting Weyl semimetal model.\footnote{There are also other holographic models for Weyl semimetal, e.g. from the pespective of fermionic spectral function \cite{Gursoy:2012ie} and top-down models \cite{Hashimoto:2016ize}. }

\para
We shall focus on the zero temperature physics. The
ansatz for the background fields at zero temperature is
\be\label{eq:bgzeroT}
ds^2=u (-dt^2+ dx^2+ dy^2)+\frac{dr^2}{u}+h dz^2\,,~~A=A_z dz\,,~~\phi=\phi(r)
\,,
\ee
where fields $u, h,A_z, \phi$ are functions of the radial coordinate $r$. Note that according to the equation of motion for $\theta$, a constant solution of $\theta$ will be found and we have set it to be zero.
The corresponding equations of motion are
\bea\label{eq:1steom}
\frac{3u''}{u}+\phi'^2-\frac{3h'u'}{2hu}-\frac{W A_z^2}{hu}&=&0\,,\\
\frac{1}{4}\phi'^2+\frac{6}{u}-\frac{3u'}{4u}\Big(\frac{u'}{u}
+\frac{h'}{h}\Big)-\frac{V}{2u}
-\frac{W A_z^2}{4uh}+\frac{Z A_z'^2}{4h}
 &=&0\,,\\
A_z''+A_z'\left(\frac{2u'}{u}-\frac{h'}{2h}+\frac{\phi' \partial_\phi Z}{Z}\right)
   -\frac{A_z W}{u Z}&=&0\,,\\
   \label{eq:4steom}
\phi''+\phi'\left(\frac{2u'}{u}+\frac{h'}{2h}\right)
   -\frac{\partial_\phi V}{u}-\frac{A_z^2 \partial_\phi W}{2h u}-\frac{A_z'^2\partial_\phi Z}{2h}&=&0\,,
\eea
where the prime is the derivative with respect to the radial coordinate $r$.
We have four independent ODEs for four unknown fields.

\para
In this paper
we will choose
\be\label{eq:modelpara}
Z(\phi)=1 \,,~~~W(\phi)=-q_0 \Big[1-\cosh\big[\sqrt{\frac{2}{3}}\phi\big]\Big]\,,~~~V(\phi)= \frac{9}{2}\Big[1-\cosh\big[\sqrt{\frac{2}{3}}\phi\big]\Big]\,.
\ee
Note that the system is invariant under the transformation $\phi\to -\phi.$ When $\phi\to 0$, we have $W(\phi)\simeq\frac{q_0}{3}\phi^2$ and $V(\phi)\simeq -\frac{3}{2}\phi^2.$ It is obvious that $q_0$ plays a similar role as axial charge and we restrict to $q_0>0$. Close to the boundary (i.e. $r\to\infty$), $\phi\to 0$, the potential in (\ref{eq:modelpara}) has the form of
$V(\phi)=\frac{1}{2}m^2 \phi^2+\dots$
with $m^2=-3$.
The coupling $Y$ does not play any role in the background solution while it plays an important role for computing the conductivities. We set
\be\label{eq:formY}
Y(\phi)=\cosh\Big[\sqrt{\frac{2}{3}}\phi \Big]\,.
\ee
\para
Close to the UV boundary we have
\be
\phi=\frac{M}{r}+\dots\,,~~~~A_z=b+\dots
\ee
and a detailed expansion will be shown in subsection \ref{subs:aefe}. $M$ and $b$ corresponds to the sources of the dual scalar operator $\bar{\psi}\psi$ and chiral current $\bar{\psi}\gamma^5\gamma^z\psi$. Turning on these two sources, the dual field theory has the same structure as the weakly coupled field theory described in appendix \ref{app:a}. In the following we shall study the bulk geometry and its free energy by tunning the parameter $M/b$ in the UV.

\subsection{Zero temperature solutions}
\label{subs:zeroT}

To study the groundstate of the system, we focus on the zero temperature solutions.\footnote{At finite temperature the ansatz of the background fields, the corresponding equations of motion and asymptotic expansions can be found in appendix \ref{app:b}.} We will first find the near horizon solutions and then turn on irrelevant perturbations to generate the full solutions. At zero temperature,  we find three different kinds of IR solutions.
\para
\noindent{\em The insulating phase.} For the insulating phase, the near horizon solution is\footnote{The Ricci scalar for the near horizon geometry at the leading order is $-3/r$ and therefore the geometry is singular near the horizon. Nevertheless the scalar potential in this solution is bounded above and satisfy the 
the Gubser criterion \cite{Gubser:2000nd, Charmousis:2010zz}. Thus the singularity is acceptable and 
the dual field theory is not pathological.}
\bea
\label{eq:in1}
u&=&r(1+r)\,,\\
\label{eq:in2}
h&=&r(1+r)\,,\\
\label{eq:in3}
A_z&=&a_1 r^{\frac{1}{4}(\sqrt{1+8q_0}-1)}\,,\\
\label{eq:in4}
\phi&=&-\sqrt{\frac{3}{2}}\log \frac{r}{1+r}\,,
\eea
where $a_1$ is a free parameter. Note that $a_1$-term is the subleading term and it sources higher oder terms in $\phi$, thus different $a_1$ will flow the geometry to different $M/b$.   The leading order of metric fields takes the form of $ds^2=r (-dt^2+dx^2+dy^2+dz^2)+\frac{dr^2}{r}$. This particular metric is known as the GPPZ  gapped geometry \cite{Girardello:1999hj} and the properties of entanglement entropy and behavior of dual scalar operators have been studied in e.g. \cite{Liu:2013una, Bianchi:2001de}. The difference is that a nontrivial $A_z$ will generate an anisotropic geometry. As we will show in subsection \ref{subsec3.1} 
there is a hard gap in the real part of diagonal optical conductivities while the anomalous Hall conductivity is nonzero at zero frequency, therefore this phase corresponds to a Chern insulator phase.
\para
\noindent{\em The Weyl semimetal phase.} The near horizon solution is
\bea\label{eq:wsm1}
u&=&r^2 \,,\\
\label{eq:wsm2}
h&=&r^2\,,\\
\label{eq:wsm3}
A_z&=&a_0+\frac{\phi_0^2}{4a_0 r}e^{-\frac{2a_0\sqrt{q_0}}{\sqrt{3}r}}\,,\\
\label{eq:wsm4}
\phi&=&\frac{\phi_0}{r^{3/2}}e^{-\frac{a_0\sqrt{q_0}}{\sqrt{3}r}}\,.
\eea
The leading order of the IR geometry is an AdS$_5$ geometry with a constant $A_z$ and we can always rescale $a_0$ to $1$. The exponential terms are the irrelevant perturbations. With the free parameter $\phi_0$, this IR geometry could flow to the whole spacetime asymptotic to AdS$_5$ with different $M/b$. This kind of near horizon also shows up in the groundstate of the holographic superconductor \cite{Horowitz:2009ij} and
the holographic Weyl semimetal phase studied in \cite{Landsteiner:2015pdh}.

\begin{figure}[h!]
\begin{center}
\includegraphics[width=0.46\textwidth]{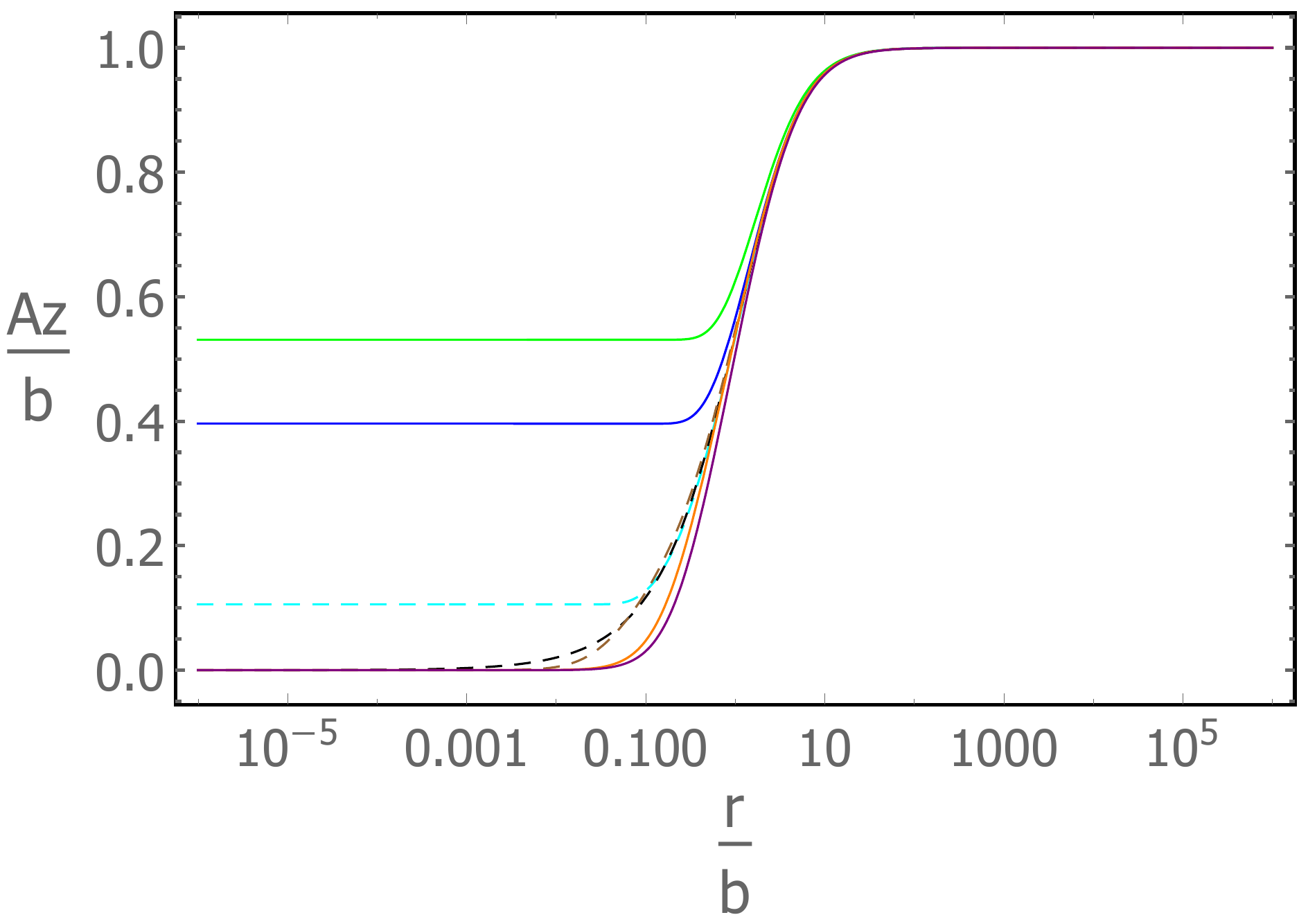}
\includegraphics[width=0.445\textwidth]{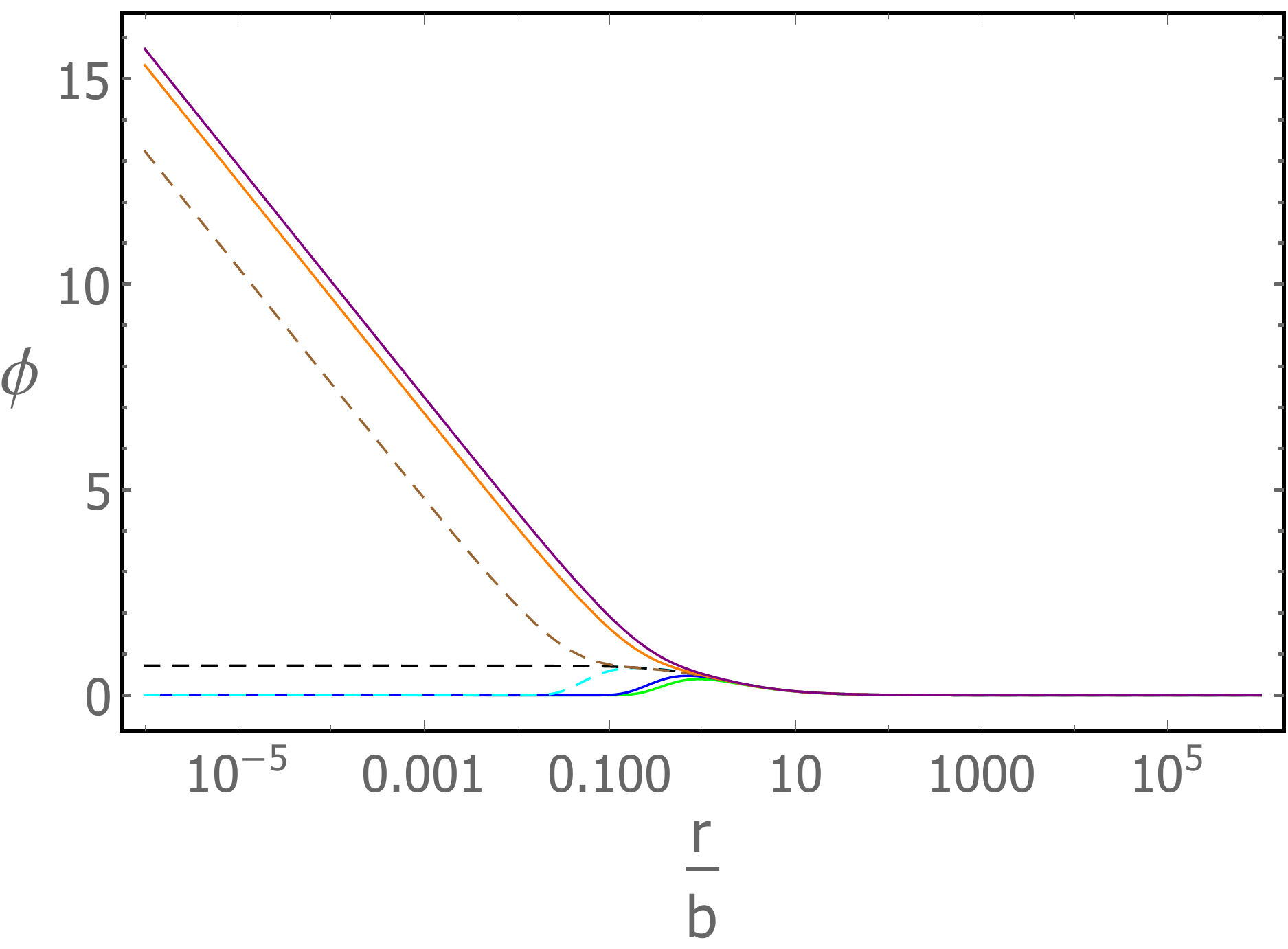}
\end{center}
\vspace{-0.6cm}
\caption{\small{The bulk profiles of $A_z$ and $\phi$ for different values of $M/b=0.941$ (green), $0.983$ (blue),
 $0.987$ (dashed cyan), $0.986$ (dashed black), $0.984$  (dashed brown), $0.987$ (orange), $1.019$ (purple). The solid lines are profiles in the stable phase while dashed lines are for the unstable phase.}}
\label{fig:bg}
\end{figure}

\para
{\em The critical point.} The near horizon solution is
\bea\label{eq:cp1}
u&=&u_0 r^2 \big(1+\delta u r^{\alpha_c} \big)\,,\\
\label{eq:cp2}
h&=&\frac{q_0}{9}r^{2\beta}\big(1+\delta h r^{\alpha_c}\big)\,,\\
\label{eq:cp3}
A_z&=&r^\beta\big(1+\delta a r^{\alpha_c} \big)\,,\\
\label{eq:cp4}
\phi&=&\sqrt{\frac{3}{2}} (\log\phi_1) \Big(1+\delta \phi r^{\alpha_c} \Big)\,.
\eea
In the case of $q_0=15$, we have $(u_0,\beta,\phi_1,\alpha_c)\simeq (1.150,0.769,1.797,1.230)$ and
$(\delta u,\delta h,\delta a)\simeq(0.147,-1.043,0.591)\delta \phi$. At the leading order there is a Lifshitz symmetry $(t, x, y, r^{-1})\to c (t, x, y, r^{-1})$, $z\to c^\beta z$ which can set $\delta\phi=-1$ to flow the Lifshitz geometry to AdS$_5$. In the UV we have $(M/b)_c\simeq0.986$. Note that for $q_0>0$ other relevant perturbations around the Lifshitz fixed point are always complex which indicates the ciritical point is unstable \cite{Hartnoll:2011pp, Donos:2012js} and we will confirm this by studying the free energy.
\para
Integrating from the above near horizon solution to the boundary, we could obtain the full solution. Different from the previous studies on holographic semimetals \cite{Landsteiner:2015pdh, Liu:2018bye}, we find that the near horizon behavior (\ref{eq:wsm1} - \ref{eq:wsm4}) flows to $M/b$ whose value runs from zero to  $(M/b)_c$, and  keeps increasing to a finite value of $(M/b)_{t+}$ with $(M/b)_{t+}>(M/b)_c$ and then turns back to $(M/b)_{c}$. While the near horizon behavior (\ref{eq:in1} - \ref{eq:in4}) flows to $M/b$ whose value runs from infinity to  $(M/b)_c$, and  keeps decreasing to a finite $(M/b)_{t-}$ with $(M/b)_{t-}<(M/b)_c$ and increases to reach $(M/b)_{c}$ finally. Examples for the bulk profiles of the matter fields at different values of $M/b$ are shown in Fig. \ref{fig:bg}. The axial gauge field and the scalar field configurations in the topological phase and the insulating phase are generally separated by the bulk profiles (dashed black) at the critical point. In the topological phase, from UV to IR the axial gauge field $A_z$ decreases monotonically and ends at a finite value in the deep IR. The scalar field is not monotonic and it first increases, then decreases to zero in the deep IR. In the insulating phase, the axial gauge field decreases from UV to zero in the IR while the scalar field increases monotonically until it hits the IR singularity. Near the critical value of $M/b$, we observe oscillatory behavior of the matter fields (dashed color lines), which is due to the complex irrelevant deformations around the Lifshtiz fixed point. This can be taken as a signature of unstable critical solution, indicating that the phase transition is not continuous, which will be confirmed from the free energy in the next subsection.

\subsection{Asymptotic expansions and free energy}
\label{subs:aefe}

\begin{figure}[h!]
\begin{center}
\includegraphics[width=0.7\textwidth]{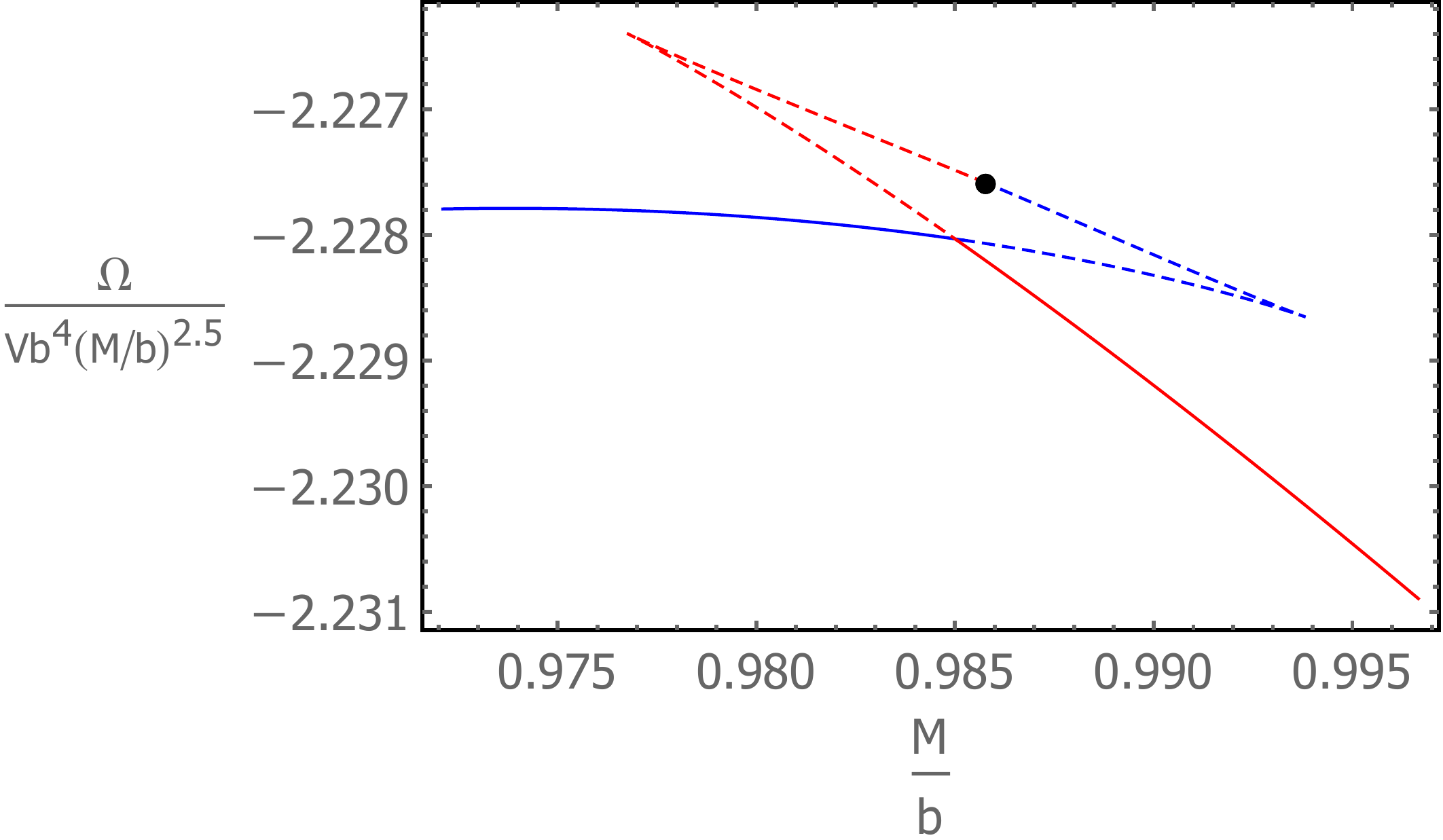}
\end{center}
\vspace{-0.7cm}
\caption{\small The free energy density as a function of $M/b$ for $q_0=15$. The blue (dashed) lines are solutions generated from the  Weyl semimetal phase while the red (dashed) line are from the insulator phase. The black dot represents the free energy at the unstable critical point. The system undergoes a first order quantum phase transition from the Weyl semimetal phase to an insulating phase. }
\label{fig:free}
\end{figure}

In order to study the stability of the background, we shall study the free energy of the bulk geometry. The asymptotic behavior and free energy for the finite temperature case can be found in \ref{app:af} and the zero temperature results can be obtained straightforwardly by setting $f=u$. At zero temperature, we have the following behaviour of fields near the UV boundary
\bea
u&=& r^2-\frac{M^2}{6}+\frac{u_2}{r^2}+...\,,\\
h&=&r^2-\frac{M^2}{6}+\frac{b^2 q_0 M^2 }{12}\frac{\log r}{r^2}+\frac{h_2}{r^2}+...\,,\\
A_z &=&b-\frac{b q_0 M^2}{6}\frac{\log r}{r^2}+\frac{\eta}{r^2}+...\,,
\eea
\bea
\phi &=&\frac{M}{r}-\frac{b^2 w_0 M }{6}\frac{\log r}{r^3}+\frac{O}{r^3}+...\,,
\eea
with
$u_2=\frac{1}{6}(b\eta-MO)+\frac{1}{72}q_0 b^2 M^2+\frac{M^4}{108}$ and $h_2=-\frac{1}{3}b\eta -\frac{1}{6}MO-\frac{1}{144}q_0 b^2 M^2+\frac{M^4}{108}.$
The free energy density can be obtained from the on-shell action to be
\be
\frac{\Omega}{V}=-\frac{1}{24} b^2 M^2 q_0 -\frac{b \eta}{2} +\frac{M^4}{48}-\frac{M O}{2}\,.
\ee
\para
With the bulk solution found in the previous subsection, we can obtain the free energy numerically.
Fig. \ref{fig:free} shows the free energy as a function of $M/b$ close to the phase transition. The critical point generated by IR geometry (\ref{eq:cp1} - \ref{eq:cp4}) is unstable and the system undergoes a first order quantum phase transition from the Weyl semimetal phase to an insulator phase.\footnote{Note that these quantum phases are not distinguished by symmetry breaking.} This behavior exists for any  $q_0>0$. Notably this is quite different from the previous holographic model \cite{Landsteiner:2015pdh} in which a continuous holographic phase transition happens between the topological Weyl semimetal phase and a trivial semimetal phase. 
The different order of the phase transition may imply the different underlying mechanics for these two types of phase transitions.

\para
The phase transitions for interacting Weyl semimetals were studied in \cite{Roy:2016rqw} from the field theoretical approach, and it was found that for sufficiently strong interactions there exists a first order quantum phase transition between the Weyl semimetal and a normal band insulator. Our holographic study  
shows that the quantum phase transition from strongly interacting Weyl semimetal to a Chern insulator (as we will show in the next section) is also of first order. Thus it broads our understanding on the phase structure of strongly interacting Weyl semimetals. 

\section{Transport properties of the dual theory}
\label{sec3}
To figure out the exact nature of the stable phases, 
we should study the conductivities. 
In the following we will compute the full frequency dependent longitudinal and transverse electric conductivities. We will also study the phase diagram from the behavior of anomalous Hall conductivity at zero frequency.
\para
The conductivities of a quantum many body system can be computed via the Kubo formula
\be
\sigma_{ij}=\lim_{\omega\to 0}\frac{1}{i\omega}\langle J_i J_j\rangle_R (\omega, {\bf k}=0)\,.
\ee
In holography, the current-current retarded correlators can be computed by studying the fluctuations of the gauge fields  dual to the currents around the background with infalling boundary conditions.

\subsection{Longitudinal conductivities}
\label{subsec3.1}
\para
We perturb the background (\ref{eq:bgzeroT}) by the fluctuation $\delta V_z=v_z(r) e^{-i\omega t}$, and obtain 
the equation 
\be\label{eq:fluvz}
v_z''+\bigg(\frac{2u'}{u}-\frac{h'}{2h}+\frac{\partial_\phi Y}{Y}\phi'\bigg)v_z'+\frac{\omega^2}{u^2}v_z=0\,.
\ee
The electric conductivities depend on the form of dilatonic coupling $Y(\phi)$ in the action (\ref{eq:action}) which is chosen to be (\ref{eq:formY}). Near the conformal boundary we have
\be
v_z=v_z^{(0)}+\frac{v_z^{(2)}}{r^2}+\frac{v_z^{(0)}\omega^2\log\Lambda r}{2r^2}+\cdots\,.
\ee
With proper boundary conditions in the IR, the optical longitudinal conductivity is then
\be
\sigma_{zz}=\frac{1}{i\omega}\bigg(2\frac{v_z^{(2)}}{v_z^{(0)}}-\frac{\omega^2}{2}\bigg)
\ee
where we have considered the counterterm to cancel the $\log \Lambda r$ term.
\para
In the phase with IR geometry (\ref{eq:in1} - \ref{eq:in4}), there are two linearly independent solutions for $v_z$ in IR  
\bea\label{eq:bc1}
v_{z1}&\simeq& c_1r^{\frac{1}{4}(1+ \sqrt{1-16\omega^2})}\Big(1+\mathcal{O}(r)\Big)\,,\\
\label{eq:bc2}
v_{z2}&\simeq& c_2
r^{\frac{1}{4}(1- \sqrt{1-16\omega^2})}\Big(1+\mathcal{O}(r)\Big)\,.
\eea
Both these two solutions are real and normalisable when $\omega<\Delta=1/4$. The unique boundary condition can be fixed by the analyticity condition of the correlator in $\omega$ \cite{Kiritsis:2015oxa}. When $\omega> \Delta$, the solutions become complex and we can choose the infalling boundary condition
\be
\label{eq:nhbc1}
v_z\simeq r^{\frac{1}{4}(1- i \sqrt{16\omega^2-1})}\Big(1+\mathcal{O}(r)\Big)\,.
\ee
The solution in (\ref{eq:bc1}) and (\ref{eq:bc2}) to produce the above infalling boundary conditon under the $\omega\to \omega+i\epsilon$ prescription is the first one, i.e. we choose the boundary condition for $v_z$ when $\omega<\Delta$ 
\be\label{eq:inphase-bc}
v_z\simeq r^{\frac{1}{4}(1+ \sqrt{1-16\omega^2})}\Big(1+\mathcal{O}(r)\Big)\,.
\ee
Since this boundary condition is real for $\omega<\Delta$, this leads to the result that the real part of the conductivity $\sigma_{zz}(\omega)$ vanishes.  There is a continuum above a gap in the optical longitudinal conductivity. Note that in this calculation we have set the unit in which the IR horizon geometry is of the form (\ref{eq:in1} - \ref{eq:in4}). As this solution flows to a specific value of $A_z\simeq b_0$ at the UV boundary, this indicates that if we set the unit $b=1$, the width of the hard gap $\Delta/b$ is  $1/4b_0$. As $b_0$ depends on the parameter $a_1$ in (\ref{eq:in3}) which will generate different $M/b$, we shall have different width of the gap for different $M/b$ in the insulating phase. Another equivalent way to see that there is indeed a hard gap in the optical conductivity is from the Schrodinger potential approach in appendix \ref{app:c}.

\begin{figure}[h!]
\begin{center}
\includegraphics[width=0.6\textwidth]{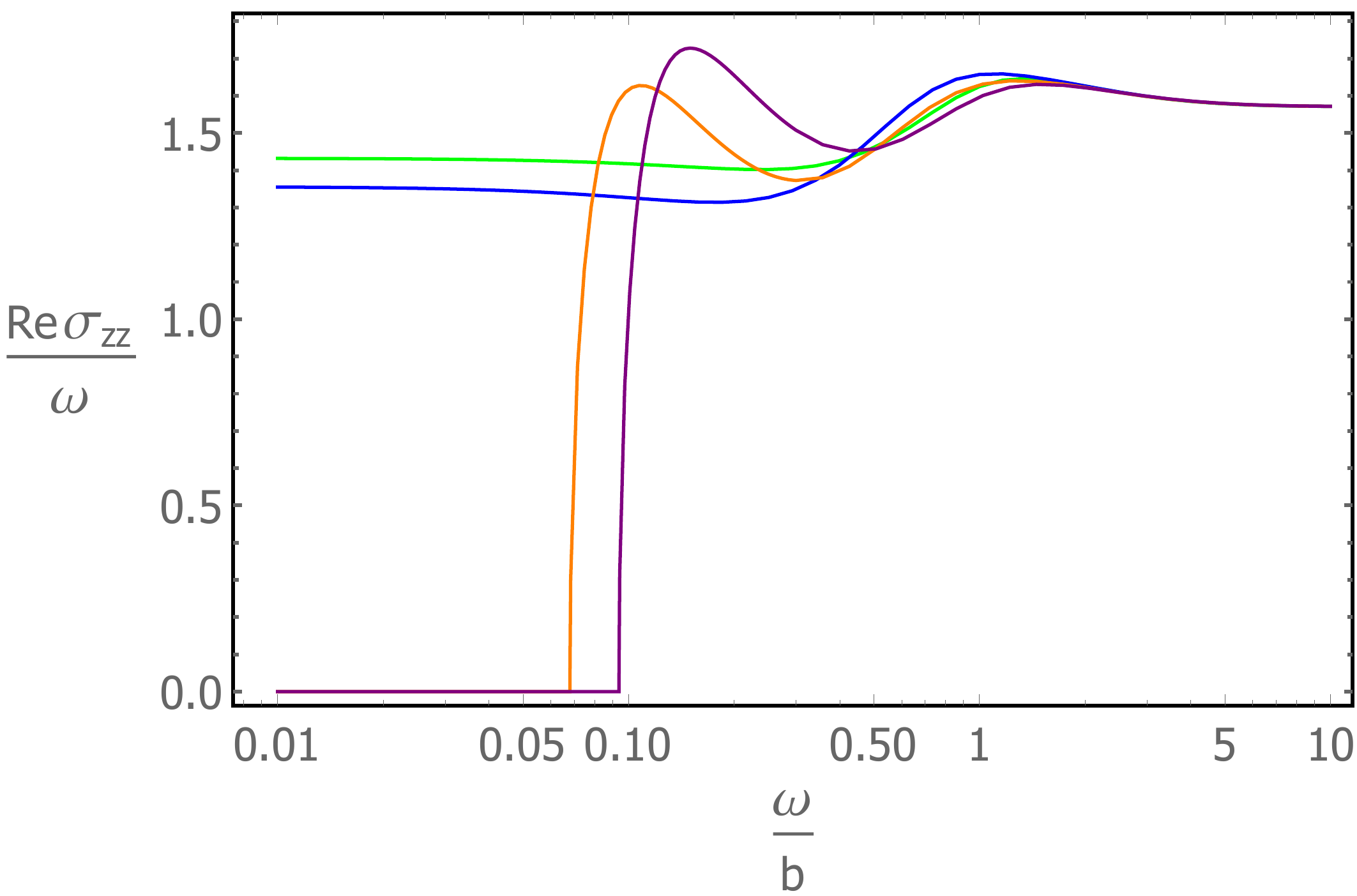}
\end{center}
\vspace{-0.5cm}
\caption{\small The real part of longitudinal conductivity as a function of the frequency $\omega/b$ for different values of $M/b=0.941$ (green), $0.983$ (blue), $0.987$ (orange), $1.019$ (purple) in the topological phase and the insulating phase.}
\label{fig:czz}
\end{figure}

\para
In the phase with IR geometry (\ref{eq:wsm1} - \ref{eq:wsm4}), the IR infalling boundary condition for $v_z$ is
\be\label{eq:nhbc2}
v_z\simeq \frac{-i\omega}{r}K_1\big[\frac{-i\omega}{r}\big]\,.
\ee
In this case there is always a continous gapless spectrum for
$\text{Re}[\sigma_{zz}(\omega)].$
\para
With the boundary conditions (\ref{eq:nhbc1}) and (\ref{eq:nhbc2}), we could obtain the retarded Green function.
In Fig. \ref{fig:czz}, we plot the real part of the longitudinal conductivity in the topological phase and the insulating phase for different values of $M/b$. In the topological phase, the longitudinal conductivity is linear in $\omega$ at both small and large frequency regimes, which is similar to the holographic results in \cite{Grignani:2016wyz}. In the insulating phase there is a hard gap in the conductivity which confirms the nature of holographic insulating phase. There exists a continuum gapless spectrum above the gap and the conductivity eventually becomes linear in $\omega$ at large frequency.

\begin{figure}[h!]
\begin{center}
\includegraphics[width=0.58\textwidth]{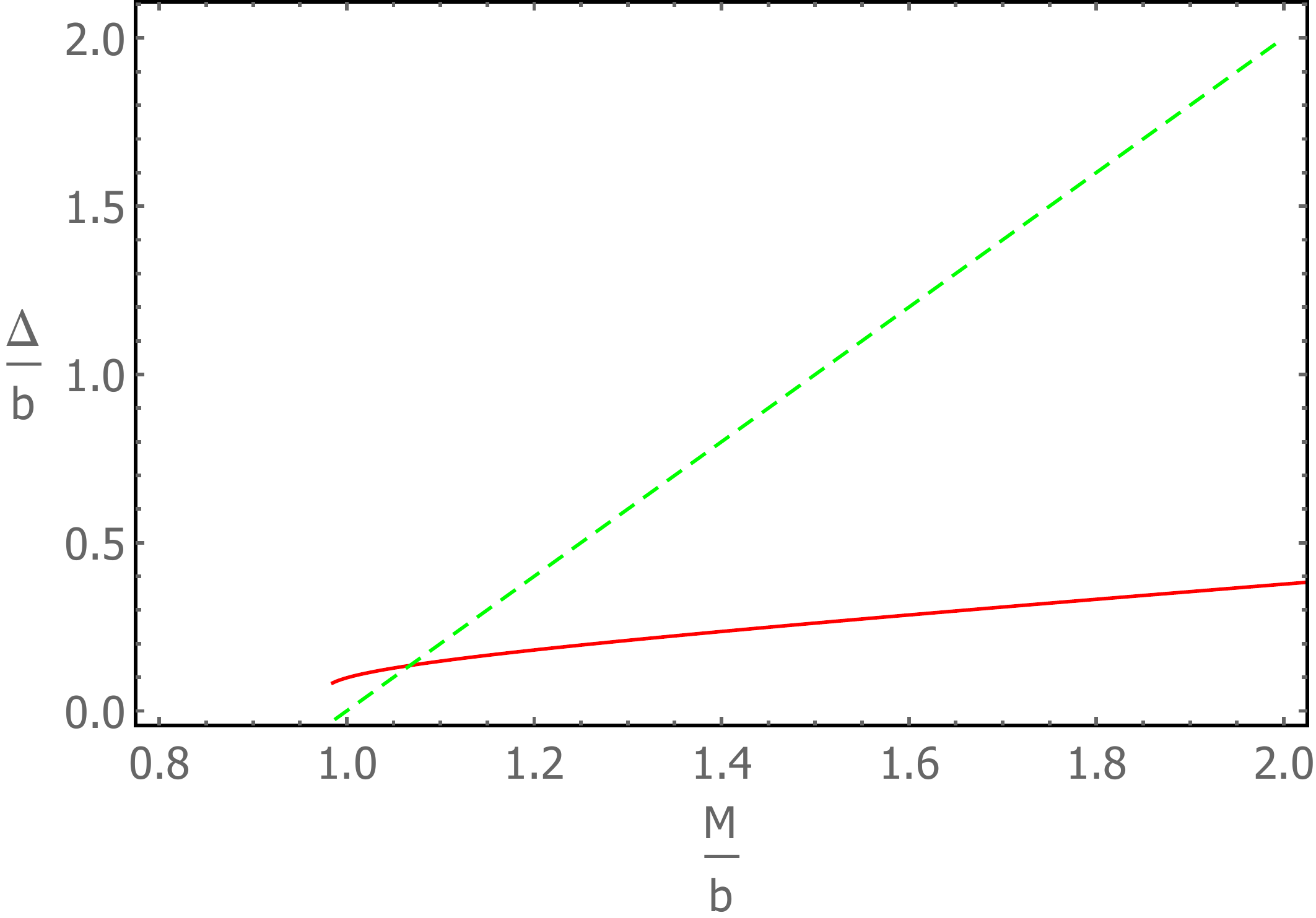}
\end{center}
\vspace{-0.5cm}
\caption{\small The dependance of the width of the hard gap as a function of $M/b$ in the holographic insulating phase (red) and weakly coupled field theory (dashed green).}
\label{fig:gap}
\end{figure}

\para
The dependance of the width of the hard gap as a function of $M/b$ is shown in Fig. \ref{fig:gap}. Similar to the weakly coupled case, it is monotonically increasing when we increase $M/b$ in the insulator phase and for large enough $M/b$, we have $\Delta/b \propto 0.22 (M/b-0.3)$.

\subsection{Transverse conductivities}
\label{subsec3.2}
The transverse conductivities can be studied by
considering  fluctuations
$\delta V_x=v_x(r) e^{-i\omega t},
\delta V_y=v_y(r) e^{-i\omega t}.$
The corresponding equations for $v_x$ and $v_y$ are
\bea\label{eq:fluvx1}
v_x''+\bigg(\frac{u'}{u}+\frac{h'}{2h}+\frac{\partial_\phi Y}{Y}\phi'\bigg)v_x'+\frac{\omega^2}{u^2}v_x+
8i\alpha\omega\frac{A_z'}{Yu\sqrt{h}}v_y &=&0\,,\\
\label{eq:fluvy1}
v_y''+\bigg(\frac{u'}{u}+\frac{h'}{2h}+\frac{\partial_\phi Y}{Y}\phi'\bigg)v_y'+\frac{\omega^2}{u^2}v_y-
8i\alpha\omega\frac{A_z'}{Yu\sqrt{h}}v_x&=&0\,.
\eea
Define $v_\pm=v_x\pm i v_y$, we obtain
\be
v_\pm''+\bigg(\frac{u'}{u}+\frac{h'}{2h}+\frac{\partial_\phi Y}{Y}\phi'\bigg)v_\pm'+\frac{\omega^2}{u^2}v_\pm \pm
8\alpha\omega\frac{A_z'}{Yu\sqrt{h}}v_\pm&=&0\,.
\ee
For the last two terms, in the deep IR $r\to 0$, the term $\frac{\omega^2}{u^2}$ always dominates. Thus the near horizon boundary conditions are the same as the case for the longitudinal conductivities. More explicitely,
for IR geometry (\ref{eq:wsm1} - \ref{eq:wsm4}), the IR infalling boundary conditions
$
v_\pm= \frac{-i\omega}{r}K_1\big[\frac{-i\omega}{r}\big],
$
while for IR geometry (\ref{eq:in1} - \ref{eq:in4}), the IR boundary conditions are
$
v_\pm\simeq r^{\frac{1}{4}(1+\sqrt{1-16\omega^2})}\big(1+\mathcal{O}(r)\big)
$
when $\omega<\Delta=1/4$ and $
v_\pm \simeq r^{\frac{1}{4}(1- i \sqrt{16\omega^2-1})}\Big(1+\mathcal{O}(r)\Big) 
$ when $\omega> \Delta$. 
\para
With the Green functions $G_\pm$ from the new variables $v_\pm$, we can compute $G_{xx}$, $G_{yy}$ and $G_{xy}$. We have $\sigma_{xy}\pm i\sigma_{xx}=\pm \frac{G_\pm}{\omega}$, i.e.  
\be\label{eq:trancond}
 \sigma_T=\sigma_{xx}=\sigma_{yy}=\frac{G_++G_-}{2i \omega}\,,~~~\sigma_{xy}=\frac{G_+-G_-}{2\omega}\,.
 \ee
The Chern-Simons term in the consistent current contributes to the anomalous Hall conductivity. We define ${\bf J}_\text{cons}=\sigma_{\text{AH}}{\bf e_b}\times {\bf E}$ and we have 
$\sigma_\text{AH}=8\alpha b-\sigma_{xy}=8\alpha b-\frac{G_+-G_-}{2\omega}$ \cite{Landsteiner:2015lsa}.

\begin{figure}[h!]
\begin{center}
\includegraphics[width=0.48\textwidth]{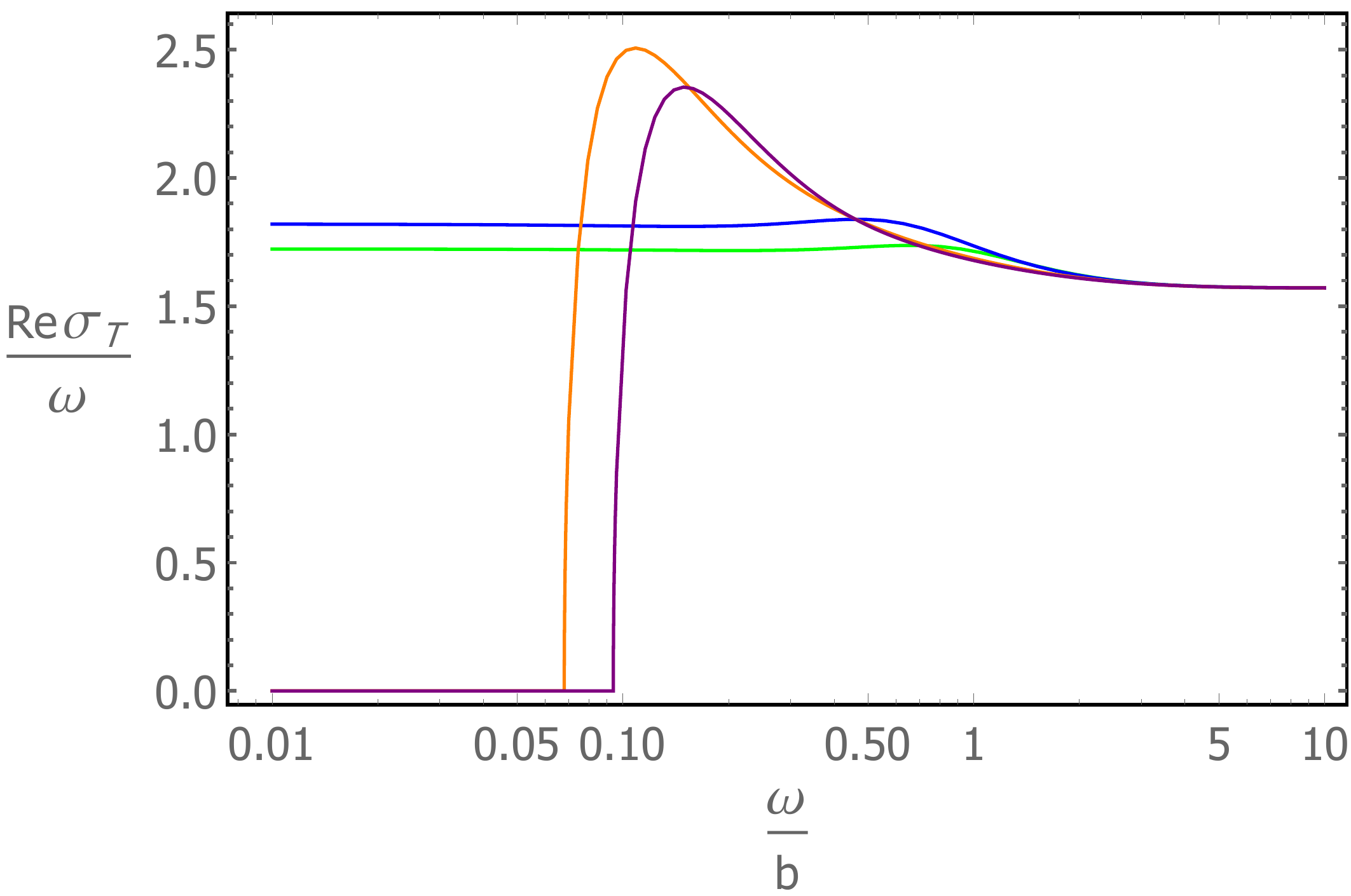}
\includegraphics[width=0.48\textwidth]{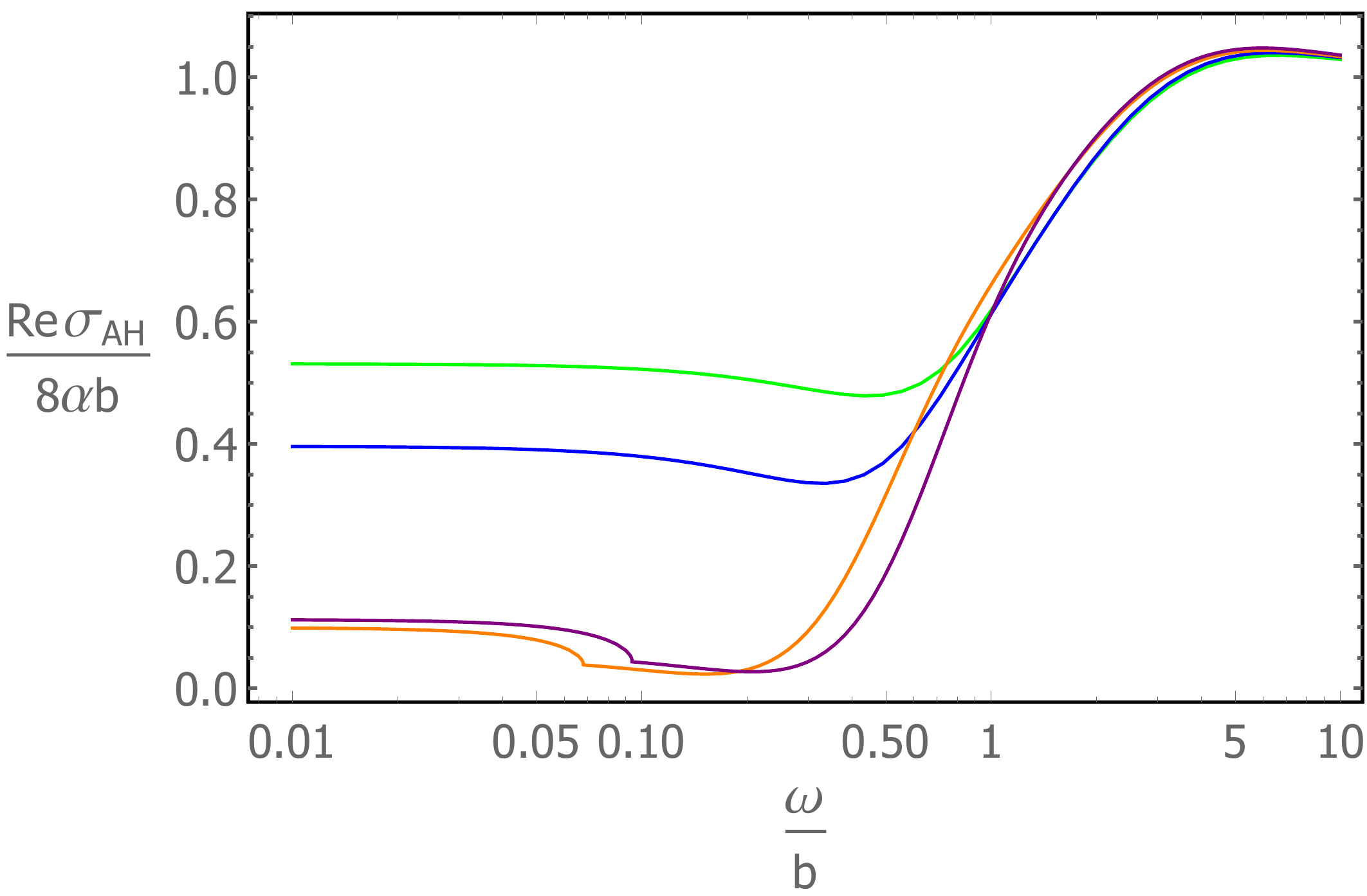}
\end{center}
\vspace{-0.6cm}
\caption{\small {The real part of the transverse conductivity (left) and the anomalous Hall conductivity (right) as a function of frequency at different values of $M/b=0.941$ (green), $0.983$ (blue), $0.987$ (orange), $1.019$ (purple) in the topological phase and the insulator phase.}}
\label{fig:cxy}
\end{figure}

\para
The numerical results for the full frequency dependent  transverse conductivities are shown in Fig. \ref{fig:cxy}.
The left plot in Fig. \ref{fig:cxy} is for the real part of optical transverse conductivities $\sigma_{xx}$ and $\sigma_{yy}$. We have a gapless spectrum in the Weyl semimetal phase. In the insulating phase, there is a continuous gapless spectrum above a hard gap $\Delta/b=1/4b_0$. This behavior is similar to the longitudinal component. 
Different from the longitudinal one, in the Weyl semimetal phase if we increase $M/b$, the ratio of the transverse conductivity $\text{Re}\sigma_T$ to the frequency increases at low frequency. This difference is caused by the emergent Lifshitz symmetry in the critical point which leads to the result that $\text{Re}[\sigma_{zz}(\omega)]\propto \omega^{2-\beta}$ while both $\text{Re}[\sigma_{T}(\omega)]$ and $\text{Re}[\sigma_\text{AH}(\omega)]$ are proportional to $\omega^{\beta}$ when $M/b$ is approaching the (unstable) critical value \cite{Landsteiner:2016stv,Grignani:2016wyz}. The right plot in Fig. \ref{fig:cxy} is the real part of optical anomalous Hall conductivity. In the insulating phase different from the diagonal component, the anomalous Hall conductivity approaches a nonzero value at zero frequency although there is an emergent time reversal symmetry in the deep IR. This is because $\sigma_\text{AH}$
depends on the real part of $G_{\pm}$, it is nonvanishing when $\omega<\Delta$ and there is no hard gap. The nonvanishing $\sigma_\text{AH}$ crucially depends on the IR boundary condition for the fluctuations which is fixed by the $\omega\to \omega+i\epsilon$ prescription. Furthermore, we observe that there is smooth change at $\omega=\Delta$ for the optical anomalous Hall conductivity in the insulating phase.  

 The behavior of conductivities in the insulating phase resembles that of a Chern insulator, indicating that our holographic model realises a quantum phase transition from a topological Weyl semimetal to a Chern insulator. In weakly coupled field  theory, there are models to describe the phase transition from a Weyl semimetal to a Chern insulator \cite{burkov1,cm-1,Roy:2016amv}. Our holographic study confirms that a similar phase structure exists for strongly interacting Weyl semimetal. 

\subsection{Phase diagram}
\label{subsec3.3}
The order parameter of the quantum phase transition between the topological phase and the insulating phase is the DC anomalous Hall conductivity. The DC anomalous Hall conductivities can be computed using a near-far matching method following \cite{Landsteiner:2015pdh}. We show here the explicit procedure of the calculations in the topological phase, and also comment on the calculations in the insulating phase.

\para
In the topological phase, near horizon the solution of $v_\pm$ with infalling boundary condition is ${v^{(n0)}_\pm}= \frac{-i\omega}{r}K_1\big[\frac{-i\omega}{r}\big].$ In the matching regime $\omega\ll r\ll \text{min} \{M, b\}$, this solution can be expanded as
\be
v^{(n0)}_\pm=1-\frac{\omega^2}{4r^2}\Big(-1+2\gamma+2\ln\Big[\frac{-i\omega}{2r}\Big]\Big)\,,
\ee
where $\gamma$ is the Euler-Mascheroni constant.
From this expansion, we know that at the matching region the infalling solution corresponds to the solution $1$ while the $\omega^2$ term can be ignored since we are interested in the $\omega\to 0$ result.
The linear order correction in $\omega$ to the near region solution is sourced by the infalling leading order solution. Thus at matching region the full linear order in $\omega$ boundary condition is
\be\label{eq:zeroTbc}
 v^{(n)}_\pm=1+\omega v^{(n1)}_\pm,
\ee where $
{v^{(n1)}_\pm}'=\mp\frac{8\alpha (A_z(r)-A_z(0))}{ r^3}\,.
$

\para
In the far region $\omega\ll r$,  we have
\be\label{eq:farreg}
{v^{(f)}_\pm}''+\Big(\frac{h'}{2h}+\frac{u'}{u}+\frac{\partial_\phi Y}{Y}\phi'\Big){v^{(f)}_\pm}'\pm\frac{8\omega\alpha }{Yu\sqrt{h}}A_z'{v^{(f)}_\pm}+\frac{\omega^2}{u^2}{v^{(f)}_\pm}=0\,.
\ee
Its solution can be expanded according to $\omega$ and we will solve the equation (\ref{eq:farreg}) up to the first order in $\omega$. Note that the last term in (\ref{eq:farreg}) can be ignored at order $\omega$. With the near horizon boundary condition (\ref{eq:zeroTbc}), we obtain the solution $v^{(f)}_\pm=1+ \omega v^{(f1)}_\pm$ where
$
{v^{(f1)}_\pm}' = \mp\frac{8\alpha (A_z(r)-A_z(0))}{Yu\sqrt{h}}.
$

\begin{figure}[h!]
\begin{center}
\includegraphics[height=6.7cm, width=0.65\textwidth]{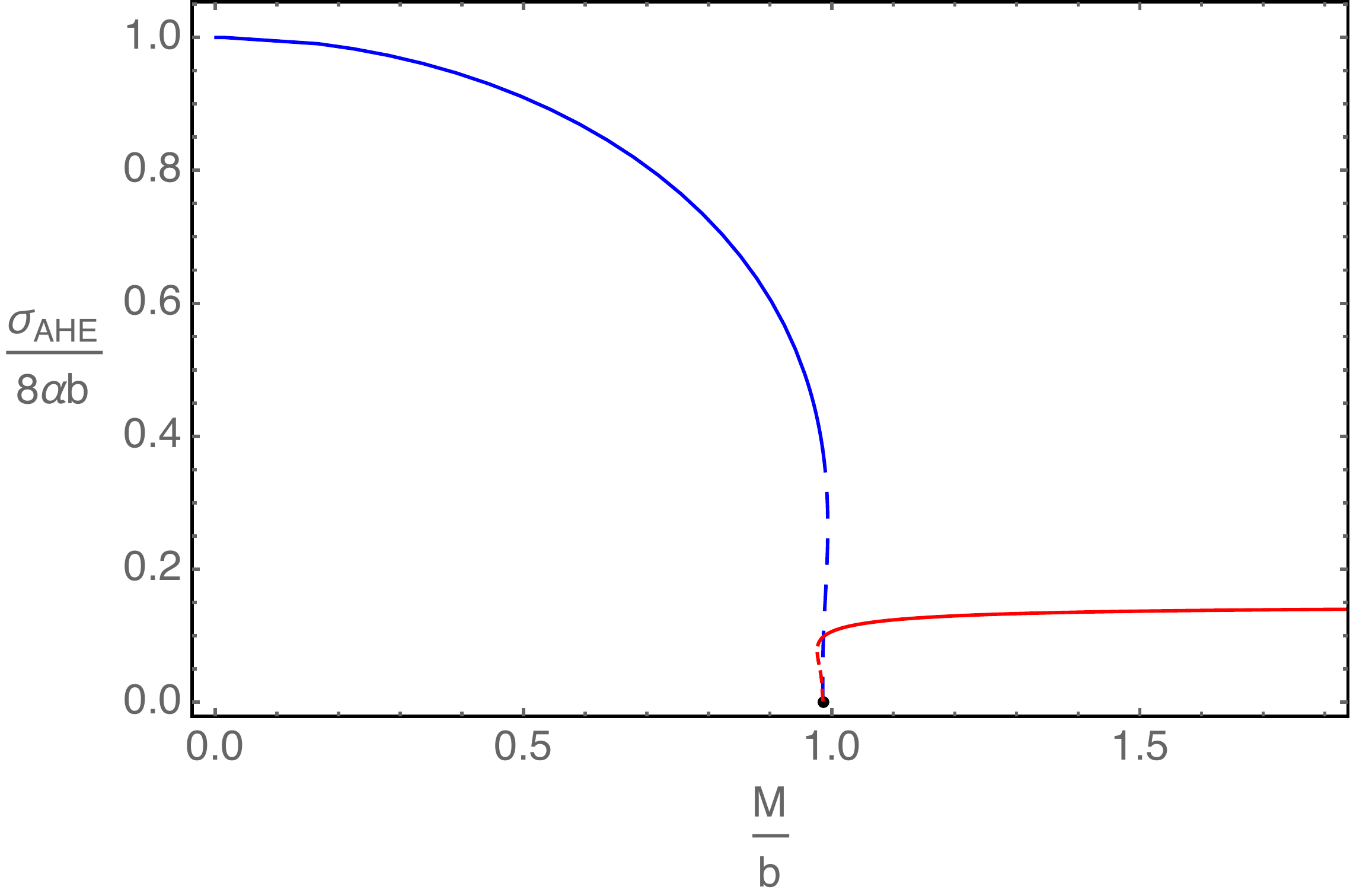}
\hspace{0.1in}
\includegraphics[height=6.6cm, width=0.3\textwidth]{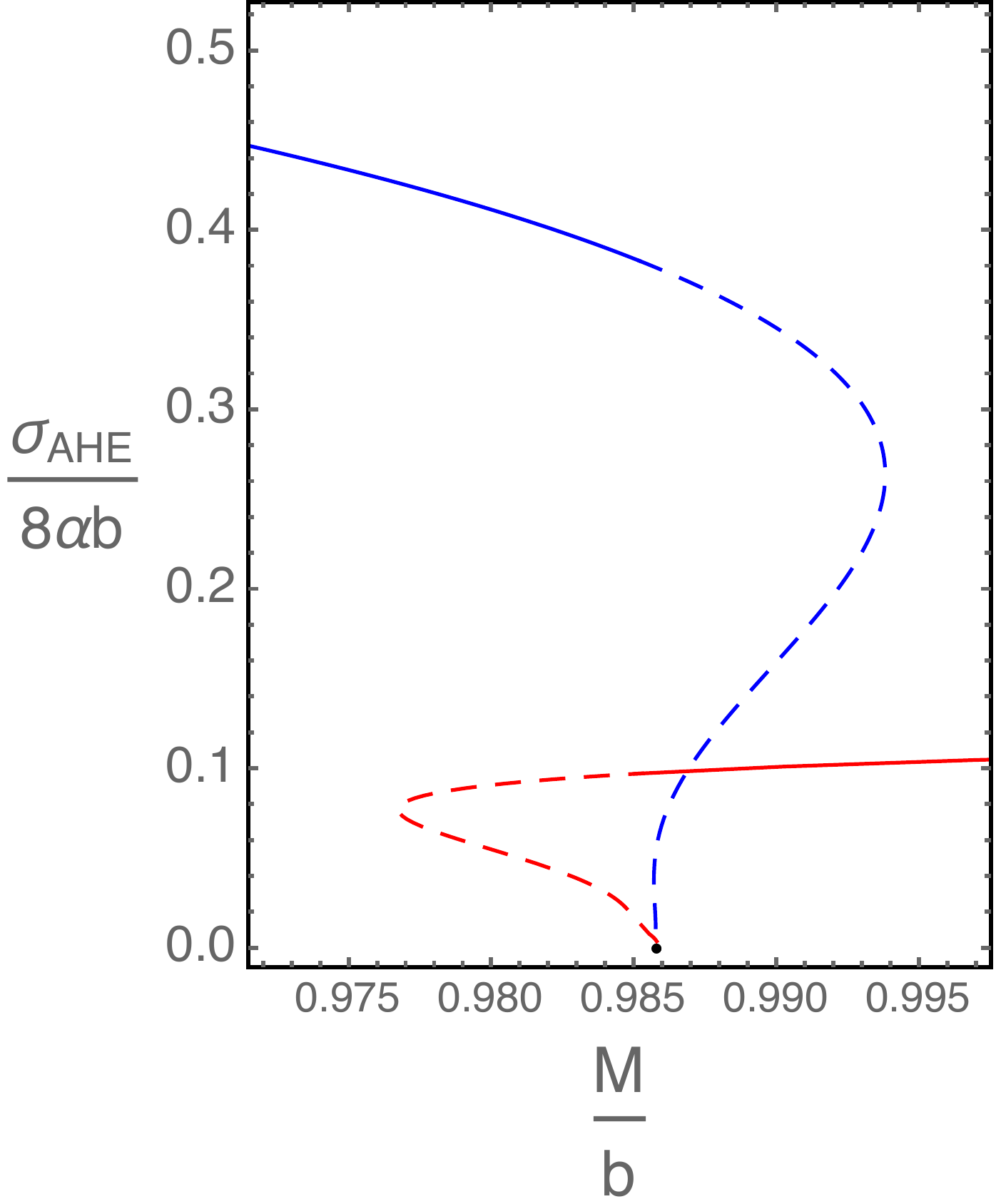}
\end{center}
\vspace{-0.6cm}
\caption{\small Both plots are for the zero temperature anomalous Hall conductivity at zero frequency of the holographic system. The right plot is a zoomed in version of the left plot close to the phase transiton point. In both plots, the blue line is for the topological Weyl semimetal phase and the red line is for the Chern insulator phase. The solid and dashed lines are for stable and unstable phases respectively. We see that under the phase transition, the anomalous Hall conductivity is discontinuous and the system undergoes a first order phase transition from a Weyl semimetal phase to a Chern insulator phase.}
\label{fig:ahe}
\end{figure}

\para
With the far region solutions, we obtain $G_\pm=\omega\big(\pm 8\alpha (b-A_z(0))\big)$. From (\ref{eq:trancond}) we obtain the DC conductivities
\be\label{eq:sigxy}
\sigma_{xy}=\frac{G_+-G_-}{2\omega}=8\alpha \big(b-A_z(0)\big)\,,~~~\sigma_{xx}=\sigma_{yy}= 0\,.
\ee
Note that in the computations above, the result of (\ref{eq:sigxy}) is for the anomalous Hall conductivity defined from the covariant currents.  In the following, we will obtain the anomalous Hall conductivity  for the consistent currents which is more close to results in real experimental systems \cite{Landsteiner:2016led, Landsteiner:2015lsa}. The final result for the zero frequency anomalous Hall conductivity in the holographic Weyl semimetal phase is
\be\label{eq:ahe}
\sigma_\text{AHE}=\text{Re}[\sigma_\text{AH}(\omega\to 0)]=8\alpha b-\sigma_{xy}=8\alpha A_z(0)
\ee
which is completely determined by the near horizon value of the axial gauge field. 
\para
In the gapped phase, one could attempt to repeat the above near-far matching method to compute the DC anomalous Hall conductivity. The solution at the matching regime should be modified to be $v^{(n)}_\pm=\sqrt{r}+\omega v^{(n1)}_\pm$, which leads to the observation that $\omega v^{(n1)}_\pm$ can not be determined analytically.   
In the insulating phase we do not have a simple formula as (\ref{eq:ahe}). This is also intuitively correct. The boundary condition (\ref{eq:inphase-bc}) is determined by the analytical continuation of the large frequency condition which reflects necessary information beyond the near horizon behavior. Therefore we have to compute the DC anomalous Hall conductivity numerically by taking $\omega\to 0$ limit of $\text{Re}[\sigma_\text{AH}(\omega)]$ obtained from the last subsection.  
\para
Fig. \ref{fig:ahe} shows the anomalous Hall conductivity at zero frequency as a function of $M/b$. The solid line is for the stable phase while the dashed line is for the unstable phase, similar to the free energy plots in Fig. \ref{fig:free}. This figure shows that when we increase $M/b$ the non-zero anomalous Hall conductivity decreases and jump directly at a phase transition point to a nonzero value which seems to be insensitive to $M/b$. 
The blue lines is for the background from IR geometry (\ref{eq:wsm1} - \ref{eq:wsm4}) while the red line is for the solutions from (\ref{eq:in1} - \ref{eq:in4}). The discontinuity of the zero frequency anomalous Hall conductivity further supports that this holographic phase transiton is of first order, which is consistent with the result from the free energy analysis. Moreover, in the holographic insulating phase, in the diagonal components of the optical conductivities there is a continuous gapless spectrum above a hard gap, and the zero frequency anomalous Hall conductivity is nonzero. These are the signals of a Chern insulator. Therefore, our holographic model describes a first order quantum phase transition from a strongly interacting Weyl semimetal to a Chern insulator. 

\section{Conclusion and discussion}
\label{sec4}
We have provided a holographic model to charaterize the quantum phase transition between the strongly interacting Weyl semimetal and the Chern insulator, by tunning the ratio between the mass parameter and time reversal symmetry breaking parameter in the dual field theory. We established that  this quantum phase transition is of first order. We also computed the frequency dependent conductivities numerically in each phase. In the holographic Weyl semimetal phase, we found that there is a nontrivial DC anomalous Hall conductivity and the diagonal components of optical conductivities are linear in frequency in  both small and large frequency regimes. In the holographic Chern insulator phase, we found that there is a hard gap in the real part of the diagonal components of the frequency dependent conductivities and there is also a nonvanishing DC anomalous Hall conductivity. This is a first example of Chern insulator from holography signified by a nontrivial anomalous Hall conductivity in a gapped state. 

\para
Our holographic model reveals the interesting phase diagram for strongly interacting Weyl semimetal and provides a novel framework to explore further problems of strongly coupled  topological states.  There are many open questions.  Firstly, in the particular holographic model we studied, the phase transition is of first order. It would be interesting to see if it is still first order for more general holographic phase transition models between Weyl semimetal and insulating phase with different dilatonic couplings. It is also important to have a better understanding of the essential physics at the first order holographic quantum phase transition point. 
Secondly, note that in field theory, there are studies of the disorder effects on the quantum phase transitions between Weyl semimetal and Chern insulators \cite{cm-1,Roy:2016amv,WSMCI}.  It would be worthwhile to explore the disorder effects or other momentum dissipation effects on this holographic quantum phase transition to understand the similarities or differences to the weakly coupled field theoretical results. Meanwhile, the transport properties of the holographic system at finite temperature is to be further explored. Finally, the insulating phase we found in this work is a Chern insulator with nontrivial anomalous Hall conductivity. It would be very interesting to study the topological invariants of this holographic Chern insulator following \cite{Liu:2018djq}, to explore effects of surface states, to realise the phase transition to a normal insulator and so on.  These studies should be helpful to build holographic models for topological insulators towards more complicated topological states of mater. We hope to explore some of these questions further in the future.

\vspace{.8cm}
\subsection*{Acknowledgments}
We thank Rong-Gen Cai,  Karl Landsteiner, Francisco Pena-Benitez, Jie Ren, Shun-Qing Shen, Ya-Wen Sun, Zhong Wang for helpful discussions. This work was  supported by the National Thousand Young Talents Program of China, NFSC Grant No.11875083 and a grant from Beihang University. Y.L. would also like to thank Hanyang University for the hospitality during the APCTP focus program ``Holography and Geometry of Quantum Entanglement'' where this work was presented.
\vspace{.3 cm}
\appendix
\section{Weakly coupled field theory for Weyl semimetal}
\label{app:a}
\begin{figure}[h!]
\begin{center}
\includegraphics[width=0.42\textwidth]{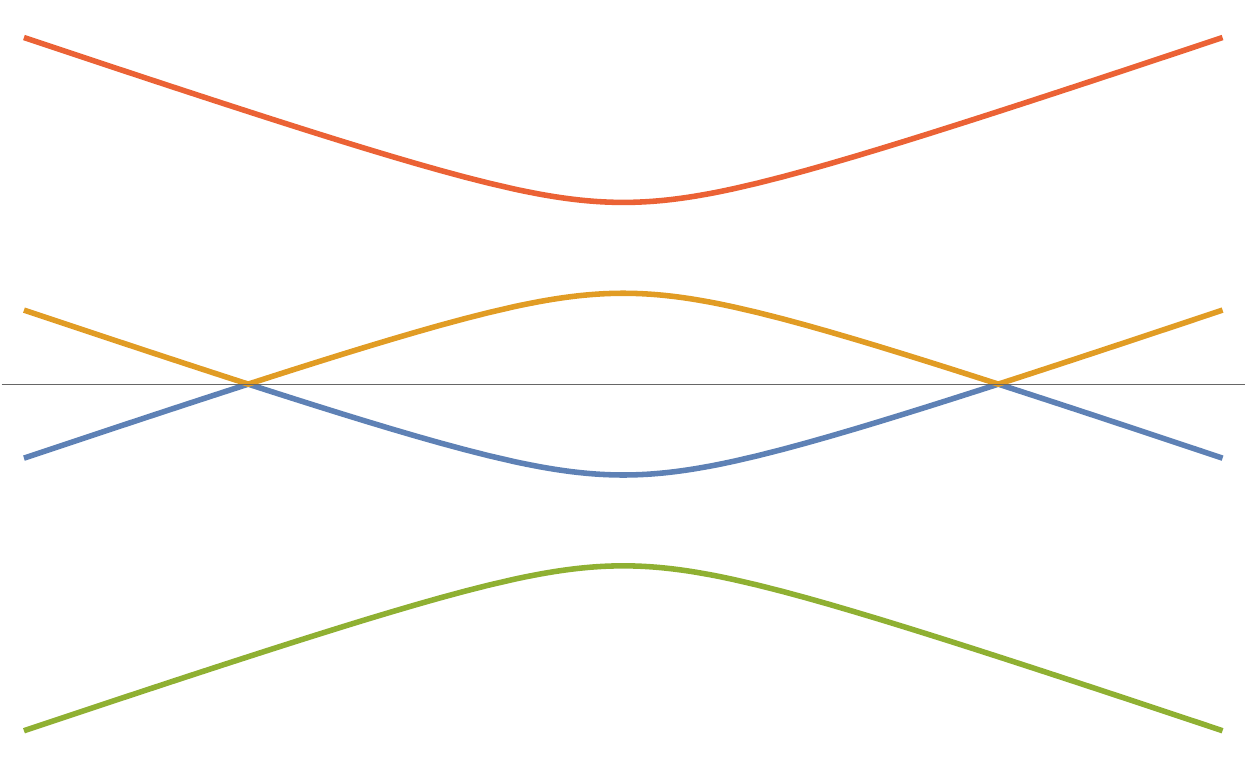}
\includegraphics[width=0.42\textwidth]{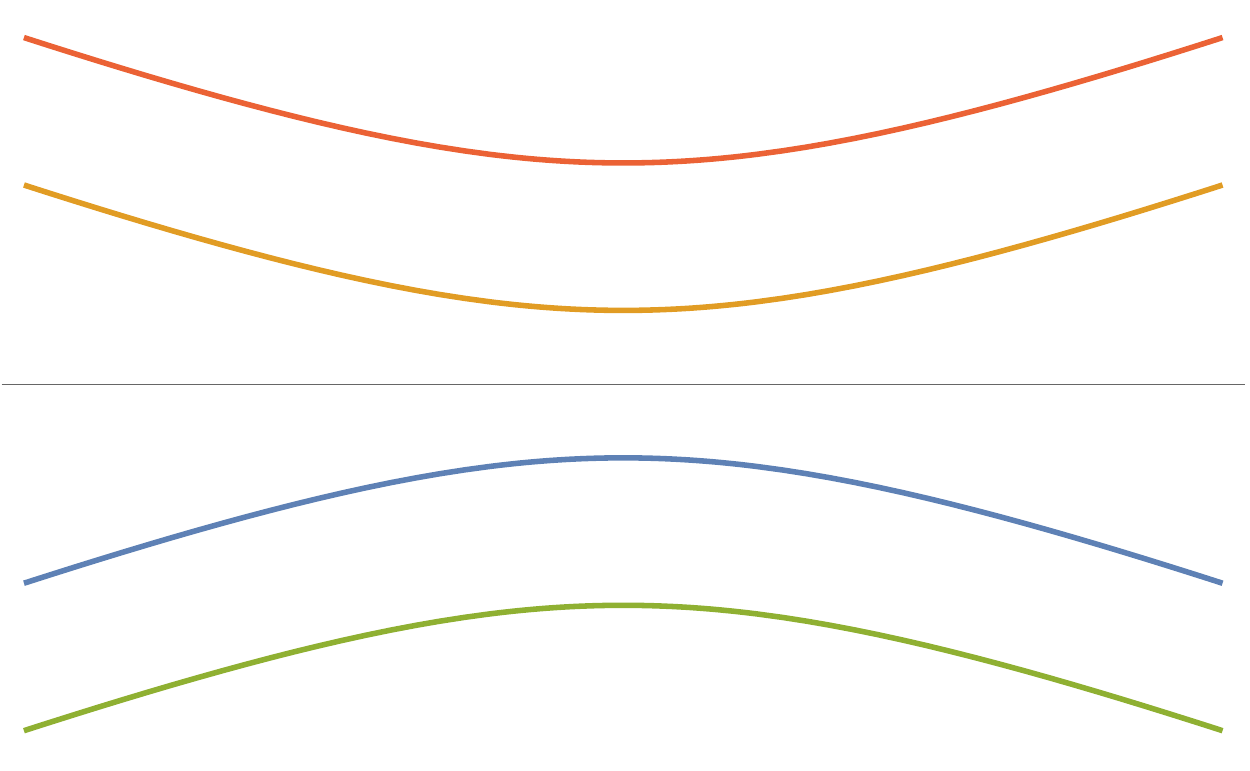}
\end{center}
\caption{\small The energy spectrum as a function of $k_z$ in the weakly coupled theory. Left: When $M<b$, there are Weyl points separating by an effective separation $2\sqrt{b^2-M^2}$ which indicates the system is in the Weyl semimetal phase; Right: When $M>b$, there is a band gap with $2 (M-b)$ in the spectrum and the system is in an insulating phase.}
\label{fig:weakly}
\end{figure}

\para
In this appendix we briefly review the weakly coupled field theory model for Weyl semimetals.
A simple low energy effective theory for Weyl semimetal \cite{Grushin:2012mt,Zyuzin:2012tv,Goswami:2012db} is
\be\label{eq:weak}
\mathcal{L}=i\bar{\psi} \big[\gamma^\mu (\partial_\mu-ie A_\mu)-\gamma_5 \gamma^\mu b_\mu \big]\psi+M\bar{\psi}\psi\,.
\ee
Here $b_0$ plays the role of axial chemical potential which breaks inversion symmetry, while $b_i$ plays the role of separation which breaks time reversal symmetry. We chose $b_\mu=b \delta_\mu^z.$ The spectrum can be computed as $E_{ij}=(-1)^i\sqrt{k_x^2+k_y^2+(b+(-1)^j\sqrt{M^2+k_z^2})^2}$ with $(i,j)=1,2.$ The energy spectrum as a function of $k_z$ while $k_x=k_y=0$ is shown in Fig. \ref{fig:weakly} from which it is clear that tunning $M/b$ the system undergoes a quantum phase transition from the Weyl semimetal to a normal band insulator. This topological phase transition can also be characterized by the DC anomalous Hall conductivity. More precisely, in the Weyl semimetal phase we have $\sigma_\text{AHE}=\frac{1}{2\pi^2}\sqrt{b^2-M^2}$ while in the insulating phase we have $\sigma_\text{AHE}=0.$ It is also interesting to note that a similar weakly coupled field theory exists for other topological semimetals, e.g. for nodal line semimetal \cite{burkov0, Liu:2018bye}.

\para
The Ward identity of the weakly coupled theory (\ref{eq:weak}) is
\bea
\nabla_\mu J^\mu=0\,,~~~\nabla_\mu J_5^\mu=\frac{1}{16\pi^2}\epsilon^{\mu\nu\rho\lambda}F_{\mu\nu}F_{\rho\lambda}+M\bar{\psi}\gamma^5\psi\,.
\eea
Notably the holographic model in the main text produced exactly the same Ward identity. 
\para
It is also possible to construct field theory model from a Weyl semimetal to a Chern insulator (3+1D anomalous Hall state) \cite{burkov1, cm-1,Roy:2016amv, WSMCI, zwang}. 
For example, if we put the above model in a lattice, by tunning $M/b$ to make the Weyl nodes located at the Brillouin zone boundary to be annihilated pairwise, the system shows a quantum phase transition from the Weyl semimetal to a Chern insulator.  In the Chern insulator phase (3+1D anomalous Hall state), the longitudinal and transverse conductivities are gapped while there is a nontrivial DC anomalous Hall conductivity. This is exactly what our holographic model realized.

\section{Equations of motion at finite temperature}
\label{app:b}
\para
In the main text we focus on the zero temperature case and for completeness we list the calculations for finite temperature in this appendix. The ansatz for the background fields at finite temperature is 
\be\label{eq:metricft}
ds^2=-udt^2+\frac{dr^2}{u}+ f(dx^2+ dy^2)+h dz^2\,,~~A=A_z dz\,,~~\phi=\phi(r)\,,
\ee
where fields $u, f, h,A_z, \phi$ are functions of the radial coordinate $r$.
\para
Plugging the ansatz (\ref{eq:metricft}) into equation, we obtain the equations of motion 
\bea\label{eq:a1steom}
\frac{f''}{f}-\frac{u''}{u}+\frac{f'h'}{2f h}-\frac{h' u'}{2h u}&=&0\,,\\
   \frac{f''}{f}+\frac{u''}{2u}-\frac{f'^2}{4f^2}
   +\frac{f'u'}{fu}-\frac{6}{u}+\frac{V}{2u}-\frac{W A_z^2}{4hu}
   -\frac{Z A_z'^2}{4h}+\frac{\phi'^2}{4}&=&0\,,\\
\frac{1}{4}\phi'^2+\frac{6}{u}-\frac{u'}{2u}\Big(\frac{f'}{f}
+\frac{h'}{2h}\Big)-\frac{f'h'}{2fh}-\frac{f'^2}{4f^2}-\frac{V}{2u}
-\frac{W A_z^2}{4uh}+\frac{Z A_z'^2}{4h}
 &=&0\,,\\
A_z''+A_z'\left(\frac{f'}{f}-\frac{h'}{2h}+\frac{u'}{u}+\frac{\phi' \partial_\phi Z}{Z}\right)
   -\frac{A_z W}{u Z}&=&0\,,\\
\phi''+\phi'\left(\frac{f'}{f}+\frac{h'}{2h}+\frac{u'}{u}\right)
   -\frac{\partial_\phi V}{u}-\frac{A_z^2 \partial_\phi W}{2h u}-\frac{A_z'^2\partial_\phi Z}{2h}&=&0\,,
\eea
where the prime is the derivative with respect to the radial coordinate $r$.
The first equation can be written as $\big(\sqrt{h}(uf'-u'f)\big)'=0$ which is essentially a conserved Noether charge.
Note that setting $f=u$, we get the equations of motion for the zero temperature case, i.e. the equation (\ref{eq:a1steom}) is trivial and we reduced to the four independent ODEs (\ref{eq:1steom} - \ref{eq:4steom}) in the main text.

\para
There is another conserved Noether charge
\be
J^r=2u^{5/2}\bigg(\frac{f\sqrt{h}}{u^{3/2}}\bigg)'+\frac{uf}{\sqrt{h}}Z A_z A_z'
\ee
satisfying $\partial_r J^r=0$ associated with the following scaling symmetry $
(x,y,z)\to c(x,y,z)\,,t\to t/c^3\,,r\to c^3 r\,,u\to c^6 u\,,
(f, h)\to(f,h)/c^2\,,A_z\to A_z/c\,,
\phi\to \phi\,.
$
This Noether charge is useful to check the accuracy of numerical code.
Meanwhile, the following scaling symmetries are useful to work in certain unit, e.g. $b=1$ for zero temperature case.\\
(I.)~~~$r \to \lambda r\,,~~(t,x,y,z) \to \lambda^{-1}(t,x,y,z)\,,~~(u,f,h) \to \lambda^2(u,f,h) \,,~~A_z \to \lambda A_z\,;$
\\ (II.)~ $(x, y) \to \lambda (x, y) \,,~~ f\to  \lambda^{-2} f\,;$\\
(III.)~$z\to \lambda z\,,~~h\to \lambda^{-2} h\,,~~A_z \to \lambda^{-1} A_z\,.$\\
We can use the first scaling symmetry to make the black hole horizon located at $r_0$ = 1 in the finite temperature case. The last two symmetry can further take the leading asymptotic coefficients of $f$ and $h$ to be 1.

\subsection{Near horizon}
\label{app:an}
\para
Near horizon $r\to r_0$, we have the expansion
\bea
u&=&4\pi T (r-r_0) +\dots\,,\\
f&=&f_1+\frac{f_1A_{z2}\big(3+10 e^{\sqrt{2/3}\phi_1}+3 e^{2\sqrt{2/3}\phi_1}\big)}{A_{z1}q_0(e^{\sqrt{2/3}\phi_1}-1)^2}(r-r_0)+\dots\,,\\
h&=&h_1+\dots\,,\\
A_z &=&A_{z1}+A_{z2} (r-r_0)+\dots\,,\\
\phi &=&\phi_1-\frac{A_{z2} \big(e^{\sqrt{2/3}\phi_1}+1\big)\big(9h_1 -A_{z1}^2 q_0\big)}{\sqrt{6} h_1 A_{z1}q_0 \big(e^{\sqrt{2/3}\phi_1}-1\big)} (r-r_0)+\dots\,,
\eea
with $T=\frac{A_{z1}q_0 e^{-\sqrt{2/3}\phi_1}(e^{\sqrt{2/3}\phi_1}-1)^2}{8\pi A_{z2}}.$ The independent parameters are $T, f_1, h_1, A_{z1}, \phi_1$. With the above scaling symmetries, we only have two free parameters, which correspond to $M/b, T/b$ in the dual field theory.

\subsection{Asymptotic behavior and free energy}
\label{app:af}

Close to the UV boundary (i.e. $r\to\infty$), 
we obtain the following behaviour of fields 
\bea
u&=& r^2-\frac{M^2}{6}+\frac{u_2}{r^2}+...\,,\\
f&=&r^2-\frac{M^2}{6}+\frac{f_2}{r^2}+...\,,\\
h&=&r^2-\frac{M^2}{6}+\frac{b^2 q_0 M^2 }{12}\frac{\log r}{r^2}+\frac{h_2}{r^2}+...\,,\\
A_z &=&b-\frac{b q_0 M^2}{6}\frac{\log r}{r^2}+\frac{\eta}{r^2}+...\,,\\
\phi &=&\frac{M}{r}-\frac{b^2 q_0 M }{6}\frac{\log r}{r^3}+\frac{O}{r^3}+...\,,
\eea
with $h_2=-2 f_2-\frac{M O}{2}+\frac{1}{48} b^2 q_0 M^2+\frac{M^4}{36}$.
Furthermore, one can obtain $f_2=u_2+\pi T  f_1 \sqrt{h_1} $ and $h_2=u_2-\frac{1}{2} b\eta-\frac{1}{48} b^2 q_0 M^2+\pi T f_1 \sqrt{h_1} $  from the two conserved Noether charges evaluated at the horizon and conformal boundary. These relations show that $u_2, f_2, h_2$ can be fully determined by $b, M, \eta, O, Tf_1\sqrt{h_1}$. Note that one can determine the above expansions only up to a shift $r\to r+a$. It is worth to point out that different from the minimal model in \cite{Landsteiner:2015pdh}, we do not have any correction of order  $\frac{\log r}{r^2}$ in $u$ and $f$.

\para
To compute the free energy, we need to obtain the on-shell action. The renormalised action is
\be
S_\text{ren}=S+S_\text{GH}+S _{\text{c.t.}}
\ee
where the Gibbons-Hawking term is $S_\text{GH}=\frac{1}{\kappa^2}\int d^4x \sqrt{-\gamma} K$ and the counterterm
\bea
S _{\text{c.t.}}&=&\frac{1}{2\kappa^2}\int d^4x \sqrt{-\gamma}\bigg[-6-\frac{\phi^2}{2}\bigg]
+\log r\int d^4x \sqrt{-\gamma}\bigg[\frac{1}{4}\mathcal{F}_{\mu\nu}\mathcal{F}^{\mu\nu}+\frac{1}{4}F_{\mu\nu}F^{\mu\nu}+
\nn\\
&&~\frac{1}{2}(\partial_\mu\phi)^2+\frac{W(\phi)}{2}A_\mu^2 \bigg]\,.
\eea
Note that $\gamma_{ab}=g_{ab}-n_a n_b$ is the induced metric on the boundary surface $r=r_\infty$ with $n^a$ the outward unit vector normal to the boundary. The trace of the extrinsic curvature is $K=\gamma^{ab}\nabla_a n_{b}$.
\para
For the ansatz (\ref{eq:metricft}), the renormalized on-shell action is
\be
\frac{S_\text{os}}{V}=\frac{1}{24} b^2 M^2 q_0 +\frac{b \eta }{2}+\pi T f_1 \sqrt{h_1} -\frac{M^4}{48}+\frac{M O}{2}\,,
\ee
therefore the free energy of the system is $\frac{\Omega}{V}=-\frac{S_\text{os}}{V}=-\frac{1}{24} b^2 M^2 q_0 -\frac{b \eta }{2}-\pi T f_1 \sqrt{h_1} +\frac{M^4}{48}-\frac{M O}{2}.$
\para
The thermodynamics of the dual system can be obtained straightforwardly. The expectation value of the stress tensor can be computed from 
\be
T_{\mu\nu}=2(K_{\mu\nu}-\gamma_{\mu\nu} K)+\frac{2}{\sqrt{-\gamma}}\frac{\delta S_\text{c.t.}}{\delta \gamma^{\mu\nu}}\,.
\ee
We obtain $\epsilon=-\frac{1}{24} b^2 M^2 q_0 -\frac{b \eta }{2}+3\pi T f_1 \sqrt{h_1} + \frac{M^4}{48}-\frac{M O}{2}$. Thus the free energy density is $\frac{\Omega}{V}=\epsilon -4\pi T f_1\sqrt{h_1}=\epsilon -Ts$ where $s$ is the entropy density expressed in unit $2\kappa^2=1$.

\section{Schrodinger potential approach to conductivities}
\label{app:c}
\para
The retarded Green's function can be computed from another equivalent approach which was widely studied in e.g. \cite{Horowitz:2009ij,Kiritsis:2015oxa}. The idea is to transform the equation of motion for the fluctuations into a Schrodinger potential problem. In the following we shall assume in the IR at the leading order the geometry is of a generic form\footnote{Obviously for the model considered in the main text, the IR geometries found in section \ref{subs:zeroT} are of this form.}
\be\label{eq:geoansatz}
u\simeq u_0 r^{\alpha_1}\,,~~ h\simeq h_0 r^{\alpha_2}\,,~~~ A_z\simeq a_1 r^{\alpha_3}\,,~~~e^\phi\simeq\phi_0 r^{\alpha_4}\,
\ee
and $Y\simeq y_0 e^{\alpha \phi}\simeq y_0 \phi_0^\alpha r^{\alpha\alpha_4}$.
 We will not explicitly solve the related Schrodinger equations, however, by analysing the behavior of Schrodinger potentials close to IR, one can conclude about which form of the leading order of geometry should one consider if the dual phase is in an insulating phase or a semimetal phase.
\para
The fluctuation equation (\ref{eq:fluvz}) can be rewritten as $
\big(\frac{u^2Y}{\sqrt{h}}v_z'\big)'+\frac{Y}{\sqrt{h}}\omega^2 v_z=0\,.
$ Introducing  $\xi$ and $\tilde{v}_z$ as
\be\label{eq:fr1}
\frac{d\xi}{dr}=\frac{1}{u}\,, ~~~\tilde{v}_z=C_1v_z\,, ~~~ C_1=\sqrt{\frac{uY}{\sqrt{h}}}\,,
\ee
we write the equation (\ref{eq:fluvz}) as a Schr\"odinger equation
\be
-\frac{d^2 \tilde{v}_z}{d\xi^2}+V_{\text{eff}}(\xi) \tilde{v}_z=\omega^2 \tilde{v}_z
\ee
with the Schr\"odinger potential
\be
V_{\text{eff}}
=\frac{(uYh^{-1/2})'}{4}\bigg(\frac{u}{Yh^{-1/2}}\bigg)'+\frac{u^2(uYh^{-1/2})''}{2uYh^{-1/2}}
\ee
where the prime is the derivative with respect to $r$. One can transfer it back to the coordinate $\xi$ using  (\ref{eq:fr1}).
\para
Depending on the values of $\alpha_1$, in the new radial variable of form (\ref{eq:fr1}), the horizon is located either at $\xi=-\infty$ (for $\alpha_1\geq 1$) or $\xi=c$  (for $\alpha_1< 1$) .
Near the UV boundary $\xi\to 0^{-}$, we have a divergent potential $V_{\text{eff}}\propto(-\xi)^{-2}$.  Near the horizon, we have 
$V_{\text{eff}}\propto r^{2\alpha_1-2}$ and its behavior will depend on $\alpha_1$.
\begin{itemize}
\item When $\alpha_1=1$, the effective potential is a constant at IR, with $V_\text{IR}=\frac{u_0^2}{4}(1+\alpha\alpha_4-\frac{\alpha_2}{2})^2$. When this constant value is positive,  the system is in the phase with a hard gap. One can check that in the model studied in the main text (\ref{eq:modelpara})
$V_\text{IR}=\frac{1}{16}$, thus we have a hard gap with width $\Delta=\frac{1}{4}$.
\item When $\alpha_1>1$, the effective potential goes to zero at IR and we have a semimetal phase.
\item When $\alpha_1<1$, the effective potential diverges and we will have a discrete spectrum for the conductivity.
\end{itemize}
\para
We can perform a similar analysis for the fluctuation equations (\ref{eq:fluvx1}, \ref{eq:fluvy1}).
Define
$
\omega_{\pm}=\omega \pm 4\alpha \frac{uA_z'}{Y\sqrt{h}}
$
we have $
\big(u\sqrt{h}Yv_\pm'\big)'+\frac{\sqrt{h}Y}{u}\omega_\pm^2 v_\pm -\frac{16u\alpha^2 A_z'^2}{\sqrt{h}Y}v_\pm=0\,.
$
We make the coordinate change and redefine the variable
\be
\frac{d\xi}{dr}=\frac{1}{u}\,,~~~~\tilde{v}_\pm=C_2 v_\pm\,,~~~ C_2= \sqrt{Yh^{1/2}}\,,
\ee
the equations (\ref{eq:fluvx1}, \ref{eq:fluvy1}) can be written as Schr\"odinger equations
\be
-\frac{d^2 \tilde{v}_\pm}{d\xi^2}+V_{\text{eff}}(\xi) \tilde{v}_z=\omega_\pm^2 \tilde{v}_\pm
\ee
with Schr\"odinger potential
\be
V_{\text{eff}}=\frac{u}{C_2}\big(uC_2'\big)'+\frac{16\alpha^2 u^2A_z'^2}{C_2^2}
=\frac{(Yh^{1/2})'}{4}\big(\frac{u^2}{Yh^{1/2}}\big)'+u^2\bigg[\frac{(Yh^{1/2})''}{2Yh^{1/2}}+\frac{16\alpha^2A_z'^2}{Y^2h}\bigg]\,.
\ee
\para
The location of the horizon in the coordinate $\xi$ is the same as the previous case, which depends on the value of $\alpha_1$ in (\ref{eq:geoansatz}).
Close to the UV boundary $\xi\to 0^{-}$, $V_{\text{eff}}\propto (-\xi)^{-2}$.
When $\alpha_1=1$, we have $V_\text{IR}=\frac{1}{16}$ and from (\ref{eq:trancond}) we know that there is a hard gap in the transport $\text{Re}[\sigma_T]$.


\vspace{.5cm}
\begin{thebibliography}{99}

\bibitem{Witten:2015aoa}
  E.~Witten,
{\em Three Lectures On Topological Phases Of Matter,}
  \doi{10.1393/ncr/i2016-10125-3}{Riv.\ Nuovo Cim.\  {\bf 39}, no. 7, 313 (2016)}
 [\arXiv{1510.07698}{cond-mat.mes-hall}].

\bibitem{vishwanath}
N.P. Armitage, E. J. Mele, A. Vishwanath
{\em Weyl and Dirac Semimetals in Three Dimensional Solids},
\doi{10.1103/RevModPhys.90.015001}{Rev. Mod. Phys. 90, 15001 (2018)}
[\arXiv{1705.01111}{cond-mat.str-el}].

\bibitem{burkov0}
A.~A.~Burkov, M.~D.~Hook and L.~ Balents,
{\em Topological nodal semimetals},
\doi{10.1103/PhysRevB.84.235126}{Phys.\ Rev.\ B {\bf 84}, 235126 (2011)}
[\arXiv{1110.1089}{cond-mat.mes-hall}].

\bibitem{Landsteiner:2016led}
  K.~Landsteiner,
{\em Notes on Anomaly Induced Transport,}
 \doi{10.5506/APhysPolB.47.2617}{Acta Phys.\ Polon.\ B {\bf 47}, 2617 (2016)}
  [\arXiv{1610.04413}{hep-th}].


\bibitem{jan}
J.~Zaanen, 
{\em Electrons go with the flow in exotic material systems,} 
\doi{10.1126/science.aaf2487}{Science 351, 1026-1027 (2016)}.


\bibitem{Zaanen:2015oix} 
  J.~Zaanen, Y.~W.~Sun, Y.~Liu and K.~Schalm,
 \href{http://www.cambridge.org/de/academic/subjects/physics/condensed-matter-physics-nanoscience-and-mesoscopic-physics/holographic-duality-condensed-matter-physics?format=HB#AlwhgydkVTSFfv7H.97}{\em Holographic Duality in Condensed Matter Physics,}  Cambridge University Press, 2015.
  \bibitem{book0}
M.~Ammon and J.~Erdmenger,
\href{http://www.cambridge.org/de/academic/subjects/physics/theoretical-physics-and-mathematical-physics/gaugegravity-duality-foundations-and-applications#xOzmEecLSr4ZJFIH.97}{\em Gauge/gravity duality: Foundations and applications},
Cambridge University Press, 2015.
\bibitem{review}
  S.~A.~Hartnoll, A.~Lucas and S.~Sachdev,
{\em Holographic quantum matter,}
[\arXiv{1612.07324}{hep-th}].

\bibitem{Landsteiner:2015pdh}
  K.~Landsteiner, Y.~Liu and Y.~W.~Sun,
{\em Quantum phase transition between a topological and a trivial semimetal from holography,}
\doi{10.1103/PhysRevLett.116.081602}{Phys.\ Rev.\ Lett.\  {\bf 116}, no. 8, 081602 (2016)}
  [\arXiv{1511.05505}{hep-th}].


\bibitem{Landsteiner:2015lsa}
  K.~Landsteiner and Y.~Liu,
 {\em The holographic Weyl semi-metal,}
 \doi{10.1016/j.physletb.2015.12.052}{Phys.\ Lett.\ B {\bf 753}, 453 (2016)}
  [\arXiv{1505.04772}{hep-th}].

\bibitem{Ammon:2016mwa}
  M.~Ammon, M.~Heinrich, A.~Jimenez-Alba and S.~Moeckel,
{\em Surface States in Holographic Weyl Semimetals,}
  \doi{10.1103/PhysRevLett.118.201601}{Phys.\ Rev.\ Lett.\  {\bf 118}, no. 20, 201601 (2017)}
  [\arXiv{1612.00836}{hep-th}].

\bibitem{Liu:2018djq} 
Y.~Liu and Y.~W.~Sun,
{\em Topological invariants for holographic semimetals,}
\arXiv{1809.00513}{hep-th}. 

\bibitem{Liu:2018bye}
  Y.~Liu and Y.~W.~Sun,
{\em Topological nodal line semimetals in holography,}
  \arXiv{1801.09357}{hep-th}.

\bibitem{Landsteiner:2016stv}
  K.~Landsteiner, Y.~Liu and Y.~W.~Sun,
 {\em Odd viscosity in the quantum critical region of a holographic Weyl semimetal,}
  \doi{10.1103/PhysRevLett.117.081604}{Phys.\ Rev.\ Lett.\  {\bf 117}, no. 8, 081604 (2016)}
  [\arXiv{1604.01346}{hep-th}].

\bibitem{Grignani:2016wyz}
  G.~Grignani, A.~Marini, F.~Pena-Benitez and S.~Speziali,
{\em AC conductivity for a holographic Weyl Semimetal,}
  \doi{10.1007/JHEP03(2017)125}{JHEP {\bf 1703}, 125 (2017)}
 [\arXiv{1612.00486}{cond-mat.str-el}].


\bibitem{Copetti:2016ewq}
  C.~Copetti, J.~Fernandez-Pendas and K.~Landsteiner,
{\em Axial Hall effect and universality of holographic Weyl semi-metals,}
  \doi{10.1007/JHEP02(2017)138}{JHEP {\bf 1702}, 138 (2017)}
  [\arXiv{1611.08125}{hep-th}].

\bibitem{Ammon:2018wzb}
  M.~Ammon, M.~Baggioli, A.~Jimenez-Alba and S.~Moeckel,
{\em A smeared quantum phase transition in disordered holography,}
 \doi{10.1007/JHEP04(2018)068}{JHEP {\bf 1804}, 068 (2018)}
  [\arXiv{1802.08650}{hep-th}].

\bibitem{Baggioli:2018afg}
  M.~Baggioli, B.~Padhi, P.~W.~Phillips and C.~Setty,
{\em Conjecture on the Butterfly Velocity across a Quantum Phase Transition,}
\doi{10.1007/JHEP07(2018)049}{JHEP {\bf 1807}, 049 (2018)} 
 [\arXiv{1805.01470}{hep-th}].

\bibitem{burkov1}
A.~A.~Burkov, L.~Balents,
{\em Weyl Semimetal in a Topological Insulator Multilayer,} 
\doi{10.1103/PhysRevLett.107.127205}{Phys. Rev. Lett. 107, 127205 (2011)} 
[\arXiv{1105.5138}{cond-mat.mes-hall}]

\bibitem{Roy:2016rqw}
  B.~Roy, P.~Goswami and V.~Juricic,
{\em Interacting Weyl fermions: Phases, phase transitions and global phase diagram,}
 \doi{10.1103/PhysRevB.95.201102}{Phys.\ Rev.\ B {\bf 95}, no. 20, 201102 (2017)}
  [\arXiv{1610.05762}{cond-mat.str-el}].

\bibitem{cm-1}
C. -Z.~ Chen, J. Song, H. Jiang, Q. -f. ~Sun, Z~ Wang, X.~ C. ~Xie,
{\em Disorder and metal-insulator transitions in Weyl semimetals,}
\doi{10.1103/PhysRevLett.115.246603}{Phys. Rev. Lett. 115, 246603 (2015)} 
[\arXiv{1507.00128}{cond-mat.mes-hall}].

\bibitem{Roy:2016amv} 
  B.~Roy, R.~J.~Slager and V.~Juricic,
{\em Global Phase Diagram of a Dirty Weyl liquid and Emergent Superuniversality,}
  [\arXiv{1610.08973}{cond-mat.mes-hall}].

\bibitem{Gursoy:2012ie}
  U.~Gursoy, V.~Jacobs, E.~Plauschinn, H.~Stoof and S.~Vandoren,
{\em Holographic models for undoped Weyl semimetals,}
  \doi{10.1007/JHEP04(2013)127}{JHEP {\bf 1304}, 127 (2013)}
  [\arXiv{1209.2593}{hep-th}].

\bibitem{Hashimoto:2016ize}
  K.~Hashimoto, S.~Kinoshita, K.~Murata and T.~Oka,
{\em Holographic Floquet states I: a strongly coupled Weyl semimetal,}
  \doi{10.1007/JHEP05(2017)127}{JHEP {\bf 1705}, 127 (2017)}
  [\arXiv{1611.03702}{hep-th}].

\bibitem{Kiritsis:2015oxa}
  E.~Kiritsis and J.~Ren,
  {\em On Holographic Insulators and Supersolids,}
\doi{10.1007/JHEP09(2015)168}{JHEP {\bf 1509}, 168 (2015)},
  [\arXiv{1503.03481}{hep-th}].

\bibitem{Gubser:2000nd}
  S.~S.~Gubser,
{\em Curvature singularities: The Good, the bad, and the naked,}
\doi{10.4310/ATMP.2000.v4.n3.a6}{Adv.\ Theor.\ Math.\ Phys.\  {\bf 4}, 679 (2000)}
  [\href{https://arxiv.org/abs/hep-th/0002160}{hep-th/0002160}].
\bibitem{Charmousis:2010zz}
  C.~Charmousis, B.~Gouteraux, B.~S.~Kim, E.~Kiritsis and R.~Meyer,
{\em Effective Holographic Theories for low-temperature condensed matter systems,}
 \doi{10.1007/JHEP11(2010)151}{JHEP {\bf 1011}, 151 (2010)}
  [\arXiv{1005.4690}{hep-th}].

\bibitem{Girardello:1999hj} 
  L.~Girardello, M.~Petrini, M.~Porrati and A.~Zaffaroni,
{\em Confinement and condensates without fine tuning in supergravity duals of gauge theories,}
  \doi{10.1088/1126-6708/1999/05/026}{JHEP {\bf 9905}, 026 (1999)} 
  [\href{https://arxiv.org/abs/hep-th/9903026}{hep-th/9903026}].

\bibitem{Liu:2013una}
  H.~Liu and M.~Mezei,
{\em Probing renormalization group flows using entanglement entropy,}
  \doi{10.1007/JHEP01(2014)098}{JHEP {\bf 1401}, 098 (2014)},
  [\arXiv{1309.6935}{hep-th}].

\bibitem{Bianchi:2001de}
  M.~Bianchi, D.~Z.~Freedman and K.~Skenderis,
 {\em How to go with an RG flow,}
  \doi{10.1088/1126-6708/2001/08/041}{JHEP {\bf 0108}, 041 (2001)}
  [\href{https://arxiv.org/abs/hep-th/0105276}{hep-th/0105276}].



\bibitem{Horowitz:2009ij}
  G.~T.~Horowitz and M.~M.~Roberts,
{\em Zero Temperature Limit of Holographic Superconductors,}
  \doi{10.1088/1126-6708/2009/11/015}{JHEP {\bf 0911}, 015 (2009)}
  [\arXiv{0908.3677}{hep-th}].

\bibitem{Hartnoll:2011pp}
  S.~A.~Hartnoll and L.~Huijse,
{\em Fractionalization of holographic Fermi surfaces,}
  \doi{10.1088/0264-9381/29/19/194001}{Class.\ Quant.\ Grav.\  {\bf 29}, 194001 (2012)}
  [\arXiv{1111.2606}{hep-th}].

\bibitem{Donos:2012js}
  A.~Donos and S.~A.~Hartnoll,
{\em Interaction-driven localization in holography,}
  \doi{10.1038/nphys2701}{Nature Phys.\  {\bf 9}, 649 (2013)}
  [\arXiv{1212.2998}{hep-th}].


\bibitem{Grushin:2012mt} 
  A.~G.~Grushin,
 {\em Consequences of a condensed matter realization of Lorentz violating QED in Weyl semi-metals,}
  \doi{10.1103/PhysRevD.86.045001}{Phys.\ Rev.\ D {\bf 86}, 045001 (2012)}
  [\arXiv{1205.3722}{hep-th}].
\bibitem{Zyuzin:2012tv} 
  A.~A.~Zyuzin and A.~A.~Burkov,
{\em Topological response in Weyl semimetals and the chiral anomaly,}
 \doi{10.1103/PhysRevB.86.115133}{Phys.\ Rev.\ B {\bf 86}, 115133 (2012)} 
  [\arXiv{1206.1868}{cond-mat.mes-hall}].
\bibitem{Goswami:2012db} 
  P.~Goswami and S.~Tewari,
{\em Axionic field theory of (3+1)-dimensional Weyl semimetals,}
\doi{10.1103/PhysRevB.88.245107}{Phys.\ Rev.\ B {\bf 88}, no. 24, 245107 (2013)} 
  [\arXiv{1210.6352}{cond-mat.mes-hall}].

\bibitem{WSMCI} 
Y. Yoshimura1, W. Onishi, K. Kobayashi, T. Ohtsuki, and K.-I. Imura, 
{\em Comparative study of Weyl semimetal, topological and Chern insulators: thin-film point of view,} 
\href{https://journals.aps.org/prb/abstract/10.1103/PhysRevB.94.235414}{Phys.Rev. B 94 (2016) 235414}, 
[\arXiv{1606.02091}{cond-mat.mes-hall}]. 


\bibitem{zwang}
L. Lu, Z. Wang, 
{\em Topological one-way fiber of second Chern number}
[\arXiv{1611.01998}{cond-mat.mes-hall}]

\end{thebibliography}
\end{document}